\documentclass[11pt]{article}

\usepackage[margin=1in]{geometry}
\usepackage{amsmath,amssymb,amsthm,mathtools}
\usepackage{enumitem}
\usepackage{booktabs}
\usepackage{array}
\usepackage{float}
\usepackage{tikz}
\usetikzlibrary{arrows.meta,positioning}
\usepackage[hidelinks]{hyperref}
\usepackage[backend=biber,style=alphabetic,maxbibnames=99]{biblatex}
\theoremstyle{plain}
\newtheorem{theorem}{Theorem}[section]
\newtheorem{lemma}[theorem]{Lemma}
\newtheorem{proposition}[theorem]{Proposition}
\newtheorem{corollary}[theorem]{Corollary}
\newtheorem*{lemma*}{Lemma}

\theoremstyle{definition}
\newtheorem*{definition*}{Definition}

\theoremstyle{definition}
\newtheorem{definition}[theorem]{Definition}

\theoremstyle{remark}
\newtheorem{remark}[theorem]{Remark}
\newtheorem*{remark*}{Remark}

\newcommand{\A}{A}
\newcommand{\Sn}{S_n}
\newcommand{\SI}{S_{\mathcal I}}
\newcommand{\abs}[1]{\left\lvert #1 \right\rvert}

\newcommand{\logn}{\log_n}
\newcommand{\gn}{\mathrm{gn}}
\newcommand{\Def}{\operatorname{Def}}
\newcommand{\Src}{\operatorname{Src}}

\title{Term Coding: An Entropic Framework for Extremal Combinatorics\\
and the Guessing-Number Sandwich Theorem}
\author{S{\o}ren Riis\\
Queen Mary University of London\\
\texttt{s.riis@qmul.ac.uk}}

\date{}

\begin{document}
\maketitle

\begin{abstract}
Classical existence problems in extremal combinatorics ask whether finite operations can satisfy prescribed
identities universally. Term Coding replaces this yes-or-no question by a graded one: for a finite system $\Gamma$,
the maximum code size $\Sn(\Gamma)$ is the largest number of satisfying assignments attainable on an $n$-element
alphabet. We prove that normalisation and diversification associate $\Gamma$ with a labelled guessing game of
guessing number $\alpha$ and give finite-alphabet sandwich bounds. Consequently,
$\log_n \Sn(\Gamma)=\alpha+o(1)$. Entropy and polymatroid inequalities provide systematic upper bounds.
Examples include a five-cycle with exponent $5/2$, self-orthogonal Latin squares, and presentation-dependent
exponents for universally equivalent identity systems.
All theorems, lemmas and propositions in this paper have been machine-checked in the Lean~4
proof assistant; the development is available at
\nolinkurl{https://github.com/SR123/term-coding-lean}.
\end{abstract}



\section{Introduction}
\label{sec:intro}

\subsection{From universal identities to graded extremal problems}

Many classical objects in extremal combinatorics are governed by \emph{universal identities} in a
finite algebraic language: Steiner triple systems via Steiner quasigroups, Latin squares via quasigroups,
and more generally designs and finite geometries via their associated incidence operations.
Traditionally one asks an \emph{existence question}: does there exist a set $\A$ of size $n$ and an
interpretation of the function symbols on $\A$ satisfying the identities for \emph{all} assignments?

The starting point of this paper is that, even when such identities cannot be satisfied everywhere,
there is a robust and informative \emph{graded} analogue.
Fix an $n$-element set $\A$ and a finite system of term equations $\Gamma$ on variables
$x_1,\dots,x_v$.
For each interpretation $\mathcal I$ of the function symbols on $\A$, let $\SI(\Gamma)\subseteq \A^v$
be the set of satisfying assignments. We then study the extremal quantity
\[
\Sn(\Gamma)\;:=\;\max_{\mathcal I}\,\abs{\SI(\Gamma)}.
\]
The equality $\Sn(\Gamma)=n^v$ holds exactly when $\Gamma$ holds identically (every assignment
satisfies every equation). Thus the classical existence question asks whether $\Sn(\Gamma)$
attains its maximal value $n^v$.
But in many natural situations $\Sn(\Gamma)$ is nontrivial and quantitatively meaningful.

For a fixed alphabet, choosing an interpretation means choosing one finite operation for each
function symbol, equivalently specifying its operation table. The same operations are then
evaluated on every assignment; they are not chosen afresh from one assignment to another.
The extremal problem is to choose these shared operations so as to maximise the number of
assignments satisfying all equations in $\Gamma$.

\subsection{A first example: Steiner triple systems and ``near misses''}
\label{subsec:sts-example}

A \emph{Steiner triple system} of order $n$ is equivalent to a \emph{Steiner quasigroup} operation
$f:\A^2\to\A$ satisfying the identities
\begin{equation}
\label{eq:sts-identities}
  f(x,x)=x,\qquad f(x,y)=f(y,x),\qquad f\bigl(x,f(x,y)\bigr)=y.
\end{equation}
It is well known that such a quasigroup exists if and only if $n\equiv 1,3\pmod 6$
(see, e.g., \cite[Ch.~19]{lintwilson2001course}).

Term Coding turns this existence theorem into a graded extremal problem.
Fix $n$ and interpret a single binary operation $f$ on $\A$.
Consider the \emph{per-pair score}
\[
  C_f\ :=\ \bigl\{(x,y)\in \A^2:\ \eqref{eq:sts-identities}\ \text{holds at }(x,y)\bigr\}.
\]
Then $\abs{C_f}$ counts how many ordered pairs satisfy all three identities simultaneously,
and
\[
  \Sn(\Gamma_{\mathrm{STS}})=\max_f\abs{C_f},
\]
where $\Gamma_{\mathrm{STS}}$ is the term-coding instance consisting of \eqref{eq:sts-identities}.
The ideal value is $n^2$, attained if and only if an STS exists.
But for values of $n$ where no STS exists, the maximum score still carries structural information
(``how close can we get?'').

\medskip
\noindent\textbf{Two concrete witnesses.}
For $n=3$ the ideal $9$ is attainable. With $\A=\{1,2,3\}$, the operation
\[
\begin{array}{c|ccc}
 f & 1 & 2 & 3\\\hline
 1 & 1 & 3 & 2\\
 2 & 3 & 2 & 1\\
 3 & 2 & 1 & 3
\end{array}
\]
satisfies \eqref{eq:sts-identities} universally.
For $n=4$ no STS exists, but one can still achieve $13$ out of $16$ pairs: with $\A=\{1,2,3,4\}$,
\[
\begin{array}{c|cccc}
 f & 1 & 2 & 3 & 4\\\hline
 1 & 1 & 1 & 1 & 1\\
 2 & 1 & 2 & 4 & 3\\
 3 & 1 & 4 & 3 & 2\\
 4 & 1 & 3 & 2 & 4
\end{array}
\]
is idempotent and commutative, and the inversion identity fails only on $(1,2),(1,3),(1,4)$.

The two witnesses already show the intended phenomenon: the score remains meaningful when the universal existence
question has a negative answer. For small alphabets, exact values of
$\Sn(\Gamma_{\mathrm{STS}})$ can be determined computationally. A direct exhaustive search,
however, ranges over all $n^{n^2}$ maps $f:\A^2\to\A$ and therefore becomes infeasible very
quickly.

\subsection{Scope}

This paper is self-contained and treats finite, single-sorted systems of equations. It develops
their entropic and graph-theoretic structure through normalisation, diversification, and
guessing numbers. Multiple sorts, disequality constraints, and the complexity of deciding
related extremal properties are outside the present scope. Section~\ref{sec:dispersion} records
how the restricted image-maximisation problem of dispersion fits into the framework.

\subsection{Relation to prior work and precise contribution}

The guessing-number framework originated in \cite{riis2007information}; for undirected graphs,
Christofides and Markstr\"om developed fractional-clique-cover bounds \cite{christofides2011guessing}, while subsequent
examples showed that these bounds can be far from sharp \cite{cameron2016guessing,baber2016graph}.
The classical guessing-game formalism assigns one local recovery rule to each vertex of a directed graph and
measures the largest set of configurations on which all local rules succeed simultaneously
\cite{riis2006information}.  Its links with information flow, network coding, and conflict-graph constructions are
developed further in \cite{gadouleau2011graph}.  The contribution here is not a new definition of the classical
guessing number.  It is a uniform reduction from \emph{arbitrary finite term-equation systems} to a source-augmented
labelled guessing game, together with finite-alphabet comparison bounds tight at the exponent level.  The labelled
interface is essential when quotienting causes several equation occurrences to define the same variable; ordered
argument positions also retain repeated and permuted inputs.

Entropy and polymatroid inequalities are established tools for bounding guessing games.  Non-Shannon inequalities
can sharpen such bounds \cite{zhang1997non,baber2016graph}.  In this paper they are used only after the syntactic
reduction, as upper-bound arguments for the resulting functional-dependence system.  The sandwich theorem itself
does not identify its optimum with an entropy-cone optimisation and assumes no general achievability statement.

The dispersion framework of \cite{riis2019max} is recovered below as a restricted image-maximisation family.
The design-theory existence results cited in the examples supply external test cases rather than ingredients of the
reduction theorem.  Thus the main result is an interface theorem: finite equational optimisation is transported to
guessing games in a form that preserves the asymptotic exponent and exposes the available graph and entropy tools.

\subsection{The computational engine: dependency graphs and guessing numbers}

Our main contribution is a two-sided graph-theoretic approximation for term coding via guessing games.
After explicit reductions, a term system gives rise to an associated directed dependency structure whose guessing-game
parameters control the maximum code size on every alphabet.
The cleanest statement is in \emph{functional normal form} (Definition~\ref{def:fnf}), where each computed variable
has a unique defining equation and the dependency structure is an ordinary digraph.
When a variable is constrained by more than one equation, we use a \emph{labelled-vertex} guessing game in which
several vertices share a common underlying variable value (equivalently, some players are known in advance to have
the same hat colour).
This yields two complementary versions of the guessing-number sandwich theorem: an unlabelled form phrased in terms
of the classical guessing game on a digraph, and a labelled form that applies to arbitrary normal-form instances
without functional completion.

At a high level, the reduction from term equations to (labelled) dependency structures proceeds via:
\begin{enumerate}
\item \textbf{Normalisation (flattening).} We introduce auxiliary variables for subterms so that every equation becomes
depth-$1$ without changing the solution set (Proposition~\ref{prop:normalisation}).
\item \textbf{Quotienting and collision elimination.} Variable equalities created by flattening can be removed by
quotienting (Lemma~\ref{lem:quotient}). Any left-hand collisions created by this identification are then eliminated by
a further quotienting step (Lemma~\ref{lem:collision-quotient}), yielding a collision-free (Term-DAG) normal form
(Definition~\ref{def:term-dag}).
\item \textbf{Diversification.} Each occurrence of a function symbol is replaced by a fresh symbol
(Definition~\ref{def:diversification}); this only rescales the alphabet size in the extremal problem
(Lemma~\ref{lem:div-sandwich}).
\item \textbf{Guessing-game equivalence.} A diversified collision-free normal-form instance is exactly a \emph{labelled}
guessing game on its labelled dependency graph (Proposition~\ref{prop:div-is-labelled-guessing}). In functional normal
form this collapses to an ordinary guessing game on the unlabelled dependency digraph
(Proposition~\ref{prop:div-is-guessing}).
\end{enumerate}
The resulting unlabelled and labelled sandwich theorems (Proposition~\ref{prop:sandwich-unlabelled} and
Theorem~\ref{thm:sandwich-labelled}) bound $\Sn(\Gamma)$ between graph parameters whenever the alphabet size
is at least the number of variables in the collision-free normal form; this threshold is asymptotically harmless.

\subsection{Non-integer exponents and the ``entropic envelope''}

A major advantage of term coding is that it naturally produces growth exponents that are \emph{not} constrained
to be integers.
This is in sharp contrast with linear-algebraic solution spaces (vector spaces) and with many classical
counting invariants.

A concrete illustration is the five-cycle $C_5$.
We will exhibit a term-coding instance whose dependency graph is the bidirected cycle on five variables.
Its guessing number is $\gn(C_5)=5/2$ \cite{riis2007graphentropy,christofides2011guessing},
so the maximum code size grows as $n^{5/2}$
(up to subpolynomial factors).
This motivates the slogan that term coding describes an \emph{entropic envelope} around universal-algebraic
constraints: the exponent is a guessing number, and the tools that bound it are polymatroid/entropy
inequalities rather than discrete rank.

\subsection{Organisation}

Section~\ref{sec:tc} defines term-coding instances and the maximum code size $\Sn(\Gamma)$.
Section~\ref{sec:normalisation} gives the normalisation procedure.
Section~\ref{sec:div} introduces diversification and the dependency graph.
Section~\ref{sec:guessing} recalls guessing games, records product constructions and convergence of guessing numbers, and
proves that diversified normal-form instances correspond to labelled guessing games; in functional normal form this collapses
to the usual guessing game on the dependency digraph.
Section~\ref{sec:sandwich} states and proves two forms of the sandwich theorem: an unlabelled version (functional case)
and a labelled version (general case).
Section~\ref{sec:entropy} explains the entropy/polymatroid viewpoint and shows that the
five-cycle entropy optimum is attained at a unique, non-integer vertex of the constraint
polytope (Proposition~\ref{prop:c5-vertex}).
Section~\ref{sec:applications} gives case studies (including $C_5$ and self-orthogonal Latin squares).
Section~\ref{sec:dispersion} briefly records dispersion as a restricted family.

\paragraph{Formal verification.}
Every theorem, lemma, proposition and corollary in this paper, together with the worked
examples of Sections~\ref{sec:applications} and~\ref{sec:dispersion}, has been formalised and
checked in the Lean~4 proof assistant (Lean 4.26.0, Mathlib v4.26.0); the development is
available at \nolinkurl{https://github.com/SR123/term-coding-lean} (release \texttt{v1.0-paper1},
archived at \nolinkurl{https://doi.org/10.5281/zenodo.22132608}). The formalisation follows
the definitions of this paper, in particular the occurrence-sensitive labelled game and the
two-layer labelled dependency graph of Definition~\ref{def:labelled-dep-graph}, and every
checked declaration depends only on the three standard axioms of Lean's core logic.
Standard entropy inequalities and the cited design-theory existence results remain external
mathematical inputs. The entropy inequalities enter only the upper bound of
Theorem~\ref{thm:c5}; the design-theory results support the citation-backed examples of
Sections~\ref{subsec:sts-example} and~\ref{subsec:sols}. The displayed normal-form
identifications are checked directly in the manuscript.

\section{Term Coding instances}
\label{sec:tc}

\subsection{Terms and equations}

Fix a finite first-order signature $\mathcal L$ consisting of function symbols $f$ of specified arities
(including $0$-ary symbols, i.e.\ constants).
Fix variables $x_1,\dots,x_v$, where throughout $v\ge 1$; the degenerate ground case with no variables can be
handled separately.
An $\mathcal L$-\emph{term} is built from the variables and function symbols in the usual way.

\begin{definition}[Term Coding instance]
\label{def:tc-instance}
A \emph{Term Coding instance} is a finite set of term equations
\[
\Gamma=\{\, s_1=t_1,\; \dots,\; s_m=t_m\,\},
\]
where each $s_i,t_i$ is an $\mathcal L$-term in the variables $x_1,\dots,x_v$.
\end{definition}

\subsection{Interpretations and codes}

Let $\A$ be a finite set of size $n$.
An \emph{interpretation} $\mathcal I$ assigns to each $k$-ary symbol $f$ a function
$f^{\mathcal I}:\A^k\to \A$, and to each constant symbol $c$ an element $c^{\mathcal I}\in \A$.
Each term $t$ then evaluates to a function $t^{\mathcal I}:\A^v\to \A$.

\begin{definition}[Code of an interpretation]
\label{def:code}
Given $\Gamma$ and an interpretation $\mathcal I$ on $\A$, define the \emph{code}
\[
\SI(\Gamma):=\bigl\{\,\mathbf a\in \A^v:\ s_i^{\mathcal I}(\mathbf a)=t_i^{\mathcal I}(\mathbf a)\ \text{for all }i=1,\dots,m\,\bigr\}.
\]
The \emph{maximum code size} over $\abs{\A}=n$ is
\[
\Sn(\Gamma):=\max_{\mathcal I}\abs{\SI(\Gamma)}.
\]
\end{definition}

\begin{remark}[Normalised exponent]
It is often convenient to work with the normalised exponent
\[
R(\Gamma,n):=\logn \Sn(\Gamma)\qquad (n\ge 2),
\]
so that $\Sn(\Gamma)=n^{R(\Gamma,n)}$ and $0\le R(\Gamma,n)\le v$.
\end{remark}

\subsection{Two guiding classes of instances}

\paragraph{Universal identities.}
If $\Gamma$ is a set of identities and we take the variables in $\Gamma$ to be precisely the universally
quantified variables, then $\Sn(\Gamma)=n^v$ if and only if there exists an algebra of size $n$ satisfying
the identities universally.
The STS instance $\Gamma_{\mathrm{STS}}$ of \eqref{eq:sts-identities} is a typical example.

\paragraph{Local constraints with global optimisation.}
More generally, the point of $\Sn(\Gamma)$ is that it is a \emph{local} score (a conjunction of local equalities)
optimised over global interpretations.
This connects naturally to local-to-global counting techniques, and, in particular, to the entropy method.
The next sections make this precise.

\section{Normalisation}
\label{sec:normalisation}

Normalisation preserves the solution set \emph{exactly} by adding auxiliary variables.
Diversification (Section~\ref{sec:div}) will not preserve the solution set exactly, but it preserves the
asymptotic exponent and makes the dependency structure transparent.

\subsection{Normal form}

\begin{definition}[Normal form]
\label{def:normal-form}
A Term Coding instance is in \emph{normal form} if every equation is of the form
\[
  f(x_{i_1},\dots,x_{i_k})=x_j\qquad\text{or}\qquad c=x_j,
\]
where $f$ is a $k$-ary function symbol and $c$ is a constant symbol.
\end{definition}

\subsection{Flattening nested terms}

\begin{proposition}[Normalisation]
\label{prop:normalisation}
For every Term Coding instance $\Gamma$ there is a normal-form instance $\Gamma^{\mathrm{nf}}$ obtained by
introducing auxiliary variables and quotienting forced variable equalities such that, for every $n$ and every
interpretation $\mathcal I$ on an $n$-element set $\A$, there is a canonical bijection
\[
\SI(\Gamma^{\mathrm{nf}})\ \longleftrightarrow\ \SI(\Gamma),
\]
and hence $\Sn(\Gamma^{\mathrm{nf}})=\Sn(\Gamma)$ for all $n$.
The forward map extends an assignment on quotient representatives to an assignment constant on each quotient class
and then restricts to the original variables; its inverse evaluates all auxiliary subterms and restricts the resulting
flat assignment to the chosen representatives.
Moreover, $\Gamma^{\mathrm{nf}}$ can be chosen to satisfy the Term-DAG/no-collision property of
Definition~\ref{def:term-dag}.
\end{proposition}

\begin{proof}
Write $V(\Gamma)$ for the variables of $\Gamma$ and let $\mathrm{Sub}(\Gamma)$ be the finite set of all \emph{syntactic subterms}
that occur in any term appearing in $\Gamma$ (including constant symbols).
For each \emph{non-variable} subterm $u\in\mathrm{Sub}(\Gamma)$ introduce a fresh variable $z_u$.
For a variable $x$ put $\widehat z_x:=x$, and for a non-variable subterm $u$ put $\widehat z_u:=z_u$.

\smallskip
\noindent\emph{Step 1: defining equations.}
For every non-variable subterm $u=f(u_1,\dots,u_k)$ (where $k$ may be $0$, in which case $f$ is a constant symbol),
add the defining equation
\[
  f(\widehat z_{u_1},\dots,\widehat z_{u_k})=z_u.
\]
\noindent\emph{Step 2: equation reduction.}
For every original equation $s=t$ of $\Gamma$, add the variable equality $\widehat z_s=\widehat z_t$.

This yields a finite system $\Gamma^{\mathrm{flat}}$ whose equations are either defining equations
$f(\cdots)=z_u$ or variable equalities.

\smallskip
\noindent\emph{Step 3: quotienting explicit equalities.}
Eliminate all variable equalities by quotienting the variable set by the equivalence relation they generate, as in
Lemma~\ref{lem:quotient}.
Denote the resulting normal-form instance by $\Gamma^{(0)}$.

\smallskip
\noindent\emph{Step 4: collision closure.}
After quotienting, distinct argument tuples may coincide and create left-hand collisions.
Apply Lemma~\ref{lem:collision-quotient} to $\Gamma^{(0)}$ to obtain an equivalent normal-form instance
$\Gamma^{\mathrm{nf}}$ that satisfies the Term-DAG/no-collision property of Definition~\ref{def:term-dag}.

It remains to compare $\Gamma^{\mathrm{flat}}$ with $\Gamma$. Fix an interpretation $\mathcal I$ on $\A$.
Let $\rho$ restrict an assignment to $V(\Gamma)$, and define
\[
  \operatorname{ext}_{\mathcal I}:\A^{V(\Gamma)}\to \A^{V(\Gamma^{\mathrm{flat}})}
\]
by retaining the original coordinates and setting
$\operatorname{ext}_{\mathcal I}(\mathbf a)(z_u)=u^{\mathcal I}(\mathbf a)$.
If $\mathbf a$ solves $\Gamma$, this extension satisfies the defining equations by construction and the equalities
$\widehat z_s=\widehat z_t$ because $s^{\mathcal I}(\mathbf a)=t^{\mathcal I}(\mathbf a)$.

Conversely, let $\mathbf b\in S_{\mathcal I}(\Gamma^{\mathrm{flat}})$ and put $\mathbf a:=\rho(\mathbf b)$.
For every non-variable subterm $u$, structural induction gives
\begin{equation}\label{eq:aux-unique}
  \mathbf b(z_u)=u^{\mathcal I}(\mathbf a).
\end{equation}
For a constant this is its defining equation. For $u=f(u_1,\dots,u_k)$, that equation gives
\[
  \mathbf b(z_u)=f^{\mathcal I}\bigl(\mathbf b(\widehat z_{u_1}),\dots,\mathbf b(\widehat z_{u_k})\bigr),
\]
where a variable argument satisfies
$\mathbf b(\widehat z_{u_i})=\mathbf a(u_i)$ and a non-variable argument is covered by the induction hypothesis.
Thus \eqref{eq:aux-unique} holds, $\mathbf b=\operatorname{ext}_{\mathcal I}(\mathbf a)$, and the equalities associated
with the original equations show that $\mathbf a$ solves $\Gamma$. Hence $\rho$ and
$\operatorname{ext}_{\mathcal I}$ are inverse on the two solution sets.

Finally, Lemmas~\ref{lem:quotient} and~\ref{lem:collision-quotient} carry this bijection through Steps 3 and 4.
Taking maxima over interpretations gives $\Sn(\Gamma^{\mathrm{nf}})=\Sn(\Gamma)$.
\end{proof}

\begin{remark}[Role of collision-freeness]
\label{rem:term-dag}
The collision-free property in Definition~\ref{def:term-dag} is precisely the output of
Step~4 of Proposition~\ref{prop:normalisation}. It is used to define the piecewise
interpretation in Lemma~\ref{lem:div-sandwich}.
\end{remark}

\begin{definition}[Term-DAG / no-collision property]
\label{def:term-dag}
Let $\Gamma$ be a normal-form instance on variables $\{x_1,\dots,x_k\}$.
We say that $\Gamma$ has the \emph{Term-DAG} (or \emph{no-collision}) property if, for each $r$-ary function symbol $f$,
distinct equations of $\Gamma$ using $f$ have distinct ordered argument tuples.
Equivalently, for every ordered $r$-tuple $(i_1,\dots,i_r)$ there is \emph{at most one} equation of the form
\[ f(x_{i_1},\dots,x_{i_r})=x_j. \]
For a $0$-ary symbol (constant) $c$, this means that $c$ occurs in at most one constant equation $c=x_j$.
\end{definition}

\begin{lemma}[Quotienting variable equalities]
\label{lem:quotient}
Let $\Gamma$ be a system of equations in normal form \emph{except} that it may additionally contain
equations of the form $x_i=x_j$ between variables.
Let $\sim$ be the equivalence relation on variables generated by these equalities, and form the quotient
instance $\Gamma/\!\sim$ by replacing each variable by a chosen representative of its $\sim$--class and then
deleting all variable-equality equations.
Then for every $n\ge 1$ and every interpretation $\mathcal I$ of the (non-variable) symbols on an alphabet
$\A$ of size $n$, restriction to representatives induces a bijection
\[
S_{\mathcal I}(\Gamma)\ \cong\ S_{\mathcal I}(\Gamma/\!\sim),
\]
and hence $\Sn(\Gamma)=\Sn(\Gamma/\!\sim)$.
\end{lemma}

\begin{proof}
Fix $\mathcal I$ and write $V$ for the set of variables of $\Gamma$.
An assignment $\mathbf a\in \A^{V}$ satisfies all variable equalities in $\Gamma$ if and only if it is
constant on each $\sim$--class.
Choose a set $R\subseteq V$ of class representatives.
Restriction gives a map
$\rho:S_{\mathcal I}(\Gamma)\to \A^R$.
Conversely, any assignment on $R$ extends uniquely to an assignment on $V$ that is constant on each
$\sim$--class.
Since $\Gamma$ and $\Gamma/\!\sim$ have the same defining equations after replacing variables by their
representatives, this extension lies in $S_{\mathcal I}(\Gamma)$ if and only if the restricted assignment
lies in $S_{\mathcal I}(\Gamma/\!\sim)$.
Thus $\rho$ is a bijection, and taking maxima over $\mathcal I$ yields the claim for $\Sn$.
\end{proof}

\begin{lemma}[Eliminating left-hand collisions]
\label{lem:collision-quotient}
Let $\Gamma$ be a normal-form instance on a finite variable set $V$.
There is an equivalent normal-form instance $\Gamma^{\dagger}$, obtained by repeatedly
identifying right-hand sides of colliding equations, such that $\Gamma^{\dagger}$ satisfies the
Term-DAG/no-collision property of Definition~\ref{def:term-dag}.

More explicitly, while $\Gamma$ contains two equations of the form
\[
  f(x_{i_1},\dots,x_{i_r})=x_j\qquad\text{and}\qquad f(x_{i_1},\dots,x_{i_r})=x_{j'}\qquad (j\neq j'),
\]
add the variable equality $x_j=x_{j'}$ and apply Lemma~\ref{lem:quotient} to quotient by the
equivalence relation generated by these added equalities.
Equations that become identical under the quotient are merged (we regard $\Gamma$ as a set of
equations; this matters in Section~\ref{sec:div}, where every equation receives its own symbol).
Since each quotient step strictly reduces the number of variables, the process terminates after finitely many steps.
The resulting normal-form instance is $\Gamma^{\dagger}$.

For every alphabet size $n\ge 1$ and every interpretation $\mathcal I$ of the symbols on an alphabet $\A$ of size $n$,
restriction to representatives induces a bijection
\[
  S_{\mathcal I}(\Gamma)\ \cong\ S_{\mathcal I}(\Gamma^{\dagger}),
\]
and hence $\Sn(\Gamma)=\Sn(\Gamma^{\dagger})$.
\end{lemma}

\begin{proof}
Each time we see a collision
$f(x_{i_1},\dots,x_{i_r})=x_j$ and $f(x_{i_1},\dots,x_{i_r})=x_{j'}$,
any satisfying assignment must have $x_j=x_{j'}$ because both sides evaluate to the same element
$f^{\mathcal I}(a_{i_1},\dots,a_{i_r})$.
Therefore adding the equality $x_j=x_{j'}$ does not change the solution set for any fixed interpretation.
Applying Lemma~\ref{lem:quotient} preserves the solution set up to canonical bijection.

Each quotient step identifies at least one pair of distinct variables, so the number of variables strictly decreases;
hence the procedure terminates.
By construction the terminal instance has no pair of distinct equations with the same left-hand side, i.e. it satisfies
Definition~\ref{def:term-dag}.
Composing the bijections from each quotient step yields the claimed bijection
$S_{\mathcal I}(\Gamma)\cong S_{\mathcal I}(\Gamma^{\dagger})$ for each $\mathcal I$.
Taking maxima over $\mathcal I$ gives $\Sn(\Gamma)=\Sn(\Gamma^{\dagger})$.
Equivalently, $\Gamma^{\dagger}=\Gamma/\!\approx$ for the least equivalence relation $\approx$
on $V$ such that whenever $f(x_{i_1},\dots,x_{i_r})=x_j$ and $f(x_{i'_1},\dots,x_{i'_r})=x_{j'}$
are equations with $x_{i_l}\approx x_{i'_l}$ for all $l$, also $x_j\approx x_{j'}$; in
particular $\Gamma^{\dagger}$ does not depend on the order in which collisions are processed.
\end{proof}

\section{Diversification and dependency graphs}
\label{sec:div}

\subsection{Sources and functional normal form}
\label{subsec:fnf}

For a normal-form instance $\Gamma$ on variables $V=\{x_1,\dots,x_k\}$, write
\[
\Def(\Gamma)\ :=\ \{x_j\in V : x_j\ \text{appears as the right-hand side of an equation in }\Gamma\}
\]
for the set of \emph{defined variables}, and
\(\Src(\Gamma):=V\setminus \Def(\Gamma)\)
for the \emph{source variables}.
Intuitively, sources are free inputs: their values are unconstrained by the equations, but they may appear
as arguments on the left-hand side of defining equations for other variables.

\begin{definition}[Functional normal form]
\label{def:fnf}
A normal-form instance $\Gamma$ is in \emph{functional normal form (FNF)} if every non-source variable is
defined by \emph{exactly one} equation, i.e.
for each $x_j\in\Def(\Gamma)$ there is a unique equation in $\Gamma$ (as a set of equations) with
right-hand side $x_j$.
\end{definition}

\begin{remark}[Scope of the guessing-game reduction]
\label{rem:scope}
For FNF instances, diversified equations give an ordinary guessing game because each non-source variable has one
local rule. Normalisation and quotienting can instead leave several equations with the same output variable.
An optional route separates those outputs with copy variables and explicit equalities;
Lemma~\ref{lem:functional-completion} gives the exact bijection. We work directly with the labelled game in
the general case (Proposition~\ref{prop:div-is-labelled-guessing} and Theorem~\ref{thm:sandwich-labelled}); the
ordinary graph remains exact in FNF and an upper bound otherwise (Propositions~\ref{prop:sandwich-unlabelled}
and~\ref{prop:div-upper}).
\end{remark}

\begin{lemma}[Functional completion]
\label{lem:functional-completion}
Let $\Gamma$ be a normal-form instance (Definition~\ref{def:normal-form}) on variables
$V=\{x_1,\dots,x_k\}$. There is an equivalent instance $\Gamma^{\mathrm{fc}}$ obtained by
splitting the outputs of the normal-form equations and recording the consistency requirements
as explicit variable equalities.

More precisely, index the normal-form equations as
$e_\ell:f_\ell(\mathbf x_\ell)=x_{j(\ell)}$. Replace the right-hand side of $e_\ell$ by a fresh
variable $x'_\ell$ and add the variable equation $x'_\ell=x_{j(\ell)}$.
Then for every alphabet size $n\ge 1$ and every interpretation $\mathcal I$ of the original
signature, the solution sets $S_{\mathcal I}(\Gamma)$ and $S_{\mathcal I}(\Gamma^{\mathrm{fc}})$
are in natural bijection (by forgetting/adding the fresh copy variables). In particular,
\[
\Sn(\Gamma^{\mathrm{fc}})=\Sn(\Gamma)\qquad\text{for all }n\ge 1.
\]
Moreover, the \emph{normal-form} part of $\Gamma^{\mathrm{fc}}$ is in functional normal form:
each computed variable is the right-hand side of at most one normal-form equation.
\end{lemma}

\begin{proof}
For fixed $\A$ and $\mathcal I$, extend a solution $\mathbf a\in \A^V$ by setting
$\widetilde{\mathbf a}(x'_\ell)=\mathbf a(x_{j(\ell)})$. The renamed equations and the added
copy equalities then hold. Conversely, those equalities turn each equation
$f_\ell(\mathbf x_\ell)=x'_\ell$ back into $f_\ell(\mathbf x_\ell)=x_{j(\ell)}$, so restriction
to $V$ is inverse to the extension. This gives the claimed bijection for every interpretation
and hence equality of the maxima. Finally, the fresh outputs $x'_\ell$ are pairwise distinct,
so the normal-form core is functional.
\end{proof}

\begin{remark*}[Copy variables as identity relays]
The copy equalities add only forced coordinates and do not change the exponent. In the guessing
game they may be viewed as identity relays, whose winning configurations correspond to those of
the quotient graph. This route is optional because the labelled game handles repeated output
variables directly.
\end{remark*}

\subsection{Diversification}

Normalisation produces flat equations but still allows the same function symbol to appear in many equations.
Diversification replaces each \emph{occurrence} of a symbol by a fresh symbol.
This yields a system where each equation is governed by its own local function.
For FNF instances this matches the guessing-game paradigm (one local rule per variable).

\begin{definition}[Diversification]
\label{def:diversification}
Let $\Gamma^{\mathrm{nf}}$ be in normal form and write its equations as
\[
  f_{\ell}(x_{i_1(\ell)},\dots,x_{i_{k_\ell}(\ell)})=x_{j(\ell)}\qquad (\ell=1,\dots,m').
\]
Here $k_\ell$ may be zero, in which case $f_\ell$ is a constant symbol.
The \emph{diversification} $\Gamma^{\mathrm{div}}$ is obtained by replacing the symbol in the $\ell$th equation
by a fresh symbol $f_\ell^{(\ell)}$ (one new symbol per equation), so that the $\ell$th equation becomes
\[
  f_\ell^{(\ell)}(x_{i_1(\ell)},\dots,x_{i_{k_\ell}(\ell)})=x_{j(\ell)}.
\]
Thus constant occurrences are diversified as fresh $0$-ary symbols as well.
\end{definition}

\subsection{Comparing original and diversified instances}

Diversification can only increase the number of solutions for a fixed alphabet size, since it gives more
freedom in interpreting symbols.
The opposite direction is also true up to a uniform scaling of the alphabet size.

\begin{lemma}[Diversification sandwich]
\label{lem:div-sandwich}
Let $\Gamma^{\mathrm{nf}}$ be a normal-form instance with the Term-DAG/no-collision property and with $k\ge 1$
variables (including any auxiliary variables introduced by normalisation), and let $\Gamma^{\mathrm{div}}$ be its
diversification.
Then for every $n\ge 1$ we have
\[
\Sn(\Gamma^{\mathrm{nf}})\ \le\ \Sn(\Gamma^{\mathrm{div}}).
\]
Moreover, if $n\ge k$ (so that $m:=\lfloor n/k\rfloor\ge 1$), then
\[
\Sn(\Gamma^{\mathrm{nf}})\ \ge\ S_{\lfloor n/k\rfloor}(\Gamma^{\mathrm{div}}).
\]
In particular, $\logn\Sn(\Gamma^{\mathrm{nf}})$ and $\logn\Sn(\Gamma^{\mathrm{div}})$ have the same $\limsup$ as
$n\to\infty$.
\end{lemma}

\begin{proof}
The first inequality is immediate: given an interpretation of $\Gamma^{\mathrm{nf}}$ on $\A$,
interpret each diversified symbol as the original symbol used in its equation.
Every solution of $\Gamma^{\mathrm{nf}}$ is then a solution of $\Gamma^{\mathrm{div}}$.

For the second inequality, assume $n\ge k$ and let $m=\lfloor n/k\rfloor$; then $m\ge 1$.
Let $\tilde\A$ be a set of size $m$.
Choose an interpretation $\tilde{\mathcal I}$ of the diversified signature on $\tilde\A$ achieving
$S_m(\Gamma^{\mathrm{div}})=\abs{S_{\tilde{\mathcal I}}(\Gamma^{\mathrm{div}})}$.

Now let $\A$ be a set of size $n$.
Partition $\A$ into $k$ disjoint blocks $\A_1,\dots,\A_k$ each of size $m$ (ignore any remaining
$n-km$ elements if $n>km$) and fix bijections $\phi_i:\tilde\A\to \A_i$.
This gives an injective block embedding
\[
  \iota:\tilde\A^k\hookrightarrow \A^k,\qquad
  \iota(\tilde a_1,\dots,\tilde a_k):=(\phi_1(\tilde a_1),\dots,\phi_k(\tilde a_k)).
\]

We define an interpretation $\mathcal I$ of the \emph{original} (non-diversified) signature on $\A$ by
simulating each diversified equation on the corresponding product block.
Consider an original $r$-ary symbol $f$ and an equation of $\Gamma^{\mathrm{nf}}$ using $f$,
\[
  f(x_{i_1},\dots,x_{i_r})=x_j,
\]
whose diversified version in $\Gamma^{\mathrm{div}}$ is
\( f^{(\ell)}(x_{i_1},\dots,x_{i_r})=x_j\).
On the product block $\A_{i_1}\times\cdots\times\A_{i_r}$ we define $f^{\mathcal I}$ by
\[
  f^{\mathcal I}(\phi_{i_1}(a_1),\dots,\phi_{i_r}(a_r))
  :=\phi_j\bigl( (f^{(\ell)})^{\tilde{\mathcal I}}(a_1,\dots,a_r)\bigr)
  \qquad (a_1,\dots,a_r\in\tilde\A).
\]
(For an equation with repeated variables, such as $f(x_1,x_1)=x_2$, the block is
$\A_1\times\A_1$ and nothing changes.)

By the Term-DAG/no-collision property (Definition~\ref{def:term-dag}), for a fixed symbol $f$ the argument tuples
$(i_1,\dots,i_r)$ appearing in equations $f(\cdots)=x_j$ of $\Gamma^{\mathrm{nf}}$ are pairwise distinct.
Consequently, if $\A_{i_1}\times\cdots\times\A_{i_r}$ and $\A_{i'_1}\times\cdots\times\A_{i'_r}$ correspond to two distinct equations,
then there exists an index $t$ with $i_t\neq i'_t$, and since the blocks $\A_1,\dots,\A_k$ are disjoint we have
$\A_{i_t}\cap\A_{i'_t}=\varnothing$, hence the two product blocks are disjoint.
Therefore the above prescriptions define a single well-defined function
$f^{\mathcal I}$ on the union of these blocks.
Outside that union, define $f^{\mathcal I}$ arbitrarily.

Finally, if a constant equation $c=x_j$ of $\Gamma^{\mathrm{nf}}$ becomes
$c^{(\ell)}=x_j$ after diversification, Definition~\ref{def:term-dag} ensures that $c$ occurs in no other constant
equation.  We then set $c^{\mathcal I}:=\phi_j((c^{(\ell)})^{\tilde{\mathcal I}})\in \A_j$.

With this interpretation, if $\tilde{\mathbf a}\in S_{\tilde{\mathcal I}}(\Gamma^{\mathrm{div}})$ then
$\mathbf a:=\iota(\tilde{\mathbf a})$ satisfies every equation of $\Gamma^{\mathrm{nf}}$:
each non-constant equation holds by the defining property of $f^{\mathcal I}$ on the corresponding product block,
and each constant equation $c=x_j$ holds because
$\mathbf a(x_j)=\phi_j(\tilde{\mathbf a}(x_j))
=\phi_j((c^{(\ell)})^{\tilde{\mathcal I}})=c^{\mathcal I}$.
Therefore $\iota$ restricts to an injection
$S_{\tilde{\mathcal I}}(\Gamma^{\mathrm{div}})\hookrightarrow S_{\mathcal I}(\Gamma^{\mathrm{nf}})$, so
\[
\abs{S_{\mathcal I}(\Gamma^{\mathrm{nf}})}
\ \ge\ \abs{S_{\tilde{\mathcal I}}(\Gamma^{\mathrm{div}})}
\ =\ S_m(\Gamma^{\mathrm{div}}).
\]
Maximising over $\mathcal I$ gives the inequality.

For the final assertion, write
\(a_n:=\log_n S_n(\Gamma^{\mathrm{nf}})\) and
\(b_n:=\log_n S_n(\Gamma^{\mathrm{div}})\) for $n\ge 2$.
The first inequality gives $\limsup a_n\le \limsup b_n$.
Conversely, taking $n=kt$ in the second inequality gives
\[
  a_{kt}\ \ge\ \frac{\log t}{\log(kt)}\,b_t .
\]
Since $\log t/\log(kt)\to 1$, taking limsups along $t\to\infty$ gives
$\limsup a_n\ge\limsup b_n$.
\end{proof}

\subsection{Dependency graphs}

\begin{definition}[Dependency digraph]
\label{def:dep-graph}
Let $\Gamma^{\mathrm{nf}}$ be a normal-form instance with variables $V=\{x_1,\dots,x_k\}$.
Its \emph{dependency digraph} is the directed graph $G_\Gamma=(V,E)$ where, for each equation
$f(x_{i_1},\dots,x_{i_r})=x_j$ in $\Gamma^{\mathrm{nf}}$, we include edges $x_{i_p}\to x_j$ for $p=1,\dots,r$.
Constant equations $c=x_j$ contribute no incoming edges.\footnote{An equation whose output variable also occurs among its arguments, such as
$f(x,y)=x$, produces a loop at $x$; loops are allowed in Definition~\ref{def:guessing-game},
and nothing below excludes this case.}

The equation nevertheless retains its ordered argument-position map
\(p\mapsto x_{i_p}\); the simple digraph records only its image and therefore forgets order and multiplicity.

Source variables are recorded separately: $x_i\in\Src(\Gamma^{\mathrm{nf}})$ iff $x_i$ is not the output of any equation
(see Subsection~\ref{subsec:fnf}).
In the associated guessing game these variables are treated as free inputs rather than as vertices that must be
guessed (Definition~\ref{def:guessing-game}).
\end{definition}

\begin{remark}[Argument positions versus in-neighbour sets]
\label{rem:argument-positions}
The forgetful passage from an ordered argument tuple to an in-neighbour set does not change
the sets of winning configurations that can be realised, and hence does not change $W_n$
(it is not a bijection of strategies: many interpretations induce the same rule).
An interpretation $f^{\mathcal I}:\A^r\to\A$ induces a rule on the distinct in-neighbours by repeating and
reordering coordinates according to $p\mapsto x_{i_p}$.  Conversely, a rule on the distinct in-neighbours defines
$f^{\mathcal I}$ on the consistent tuples---those in which positions carrying the same variable have equal
values---and may be extended arbitrarily to the remaining tuples.  All guessing-game identifications below use
these two maps.  In a formal treatment the argument-position map is primary and the simple dependency digraph is
its forgetful image.
The same observation applies when several in-neighbours of a vertex carry the same label in
Definition~\ref{def:labelled-guessing-game}: the local function then receives the same value
more than once.
\end{remark}

The same graph is associated to $\Gamma^{\mathrm{div}}$ since diversification does not change which variables
occur as inputs/outputs, only which symbol names them.

\subsection{A visual example: the five-cycle \texorpdfstring{$C_5$}{C5}}
\label{subsec:c5-visual}

Normalisation and diversification are easiest to digest with an example.
Consider a single binary symbol $f$ and variables $x,y,z$ with the three nested equations
\begin{equation}
\label{eq:c5-nested}
  f(f(z,x),y)=x,\qquad f(x,f(y,z))=y,\qquad f(f(y,z),f(z,x))=z.
\end{equation}
This is a compact presentation, but the dependency structure is hidden by nested terms.

\paragraph{Normalisation.}
Introduce auxiliary variables
\[
  \alpha := f(z,x),\qquad \beta := f(y,z).
\]
Then \eqref{eq:c5-nested} is equivalent (Proposition~\ref{prop:normalisation}) to the normal form system
\begin{equation}
\label{eq:c5-nf}
  f(z,x)=\alpha,\qquad f(y,z)=\beta,\qquad f(\alpha,y)=x,\qquad f(x,\beta)=y,\qquad f(\beta,\alpha)=z.
\end{equation}

\paragraph{Diversification.}
Replace the five occurrences of $f$ by fresh symbols $f_1,\dots,f_5$.
The diversified instance becomes
\begin{equation}
\label{eq:c5-div}
  f_1(z,x)=\alpha,\quad f_2(y,z)=\beta,\quad f_3(\alpha,y)=x,\quad f_4(x,\beta)=y,\quad f_5(\beta,\alpha)=z.
\end{equation}

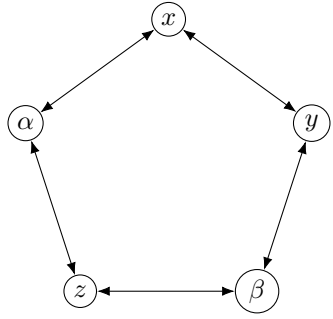
\begin{figure}[H]
\centering
\begin{tikzpicture}[>=Latex, every node/.style={draw,circle,inner sep=2pt,font=\small}]
  \node (x)     at (90:2)   {$x$};
  \node (y)     at (18:2)   {$y$};
  \node (b)     at (306:2)  {$\beta$};
  \node (z)     at (234:2)  {$z$};
  \node (a)     at (162:2)  {$\alpha$};
  \path[<->]
    (x) edge (y)
    (y) edge (b)
    (b) edge (z)
    (z) edge (a)
    (a) edge (x);
\end{tikzpicture}
\caption{The dependency graph of the diversified $C_5$ instance \eqref{eq:c5-div}: the bidirected 5-cycle on
$\{x,y,z,\alpha,\beta\}$.}
\label{fig:c5}
\end{figure}

\paragraph{The dependency graph.}
The variables are $\{x,y,z,\alpha,\beta\}$.
Reading off edges from \eqref{eq:c5-div} shows that the underlying undirected graph is the cycle
\[x\!\leftrightarrow\!y\!\leftrightarrow\!\beta\!\leftrightarrow\!z\!\leftrightarrow\!\alpha\!\leftrightarrow\!x,\]
with each edge bidirected.
Figure~\ref{fig:c5} depicts this graph.

We will return to this example in Section~\ref{subsec:c5-exponent} and explain why its exponent is $5/2$.

\section{Guessing games and diversified instances}
\label{sec:guessing}

\subsection{Guessing games}

We recall Riis' guessing-game formalism \cite{riis2006information} in a form adapted to term coding.
Two mild extensions are convenient for us:
(i) we allow a distinguished set of \emph{source} vertices, modelling term-theoretic sources (variables not
defined by any equation), and
(ii) we allow a \emph{labelled-vertex} variant in which several vertices share the same underlying variable value
(useful when a variable is constrained by more than one equation).

\begin{definition}[Guessing game with sources]
\label{def:guessing-game}
Let $G=(V,E)$ be a finite directed graph (loops allowed) and let $S\subseteq V$ be a designated set of
\emph{source vertices}. Fix an alphabet $\A$ with $|\A|=n$.

A (deterministic) \emph{strategy} on $(G,S)$ is a family of local functions
\[
  g_v:\A^{N^-(v)}\to \A \qquad (v\in V\setminus S),
\]
where $N^-(v)$ denotes the set of in-neighbours of $v$.
A configuration $\mathbf a=(a_v)_{v\in V}\in\A^V$ is \emph{winning} for the strategy if
\[
  a_v\ =\ g_v\bigl((a_u)_{u\in N^-(v)}\bigr)\qquad\text{for all }v\in V\setminus S.
\]
(No condition is imposed at vertices in $S$.)

Let $W_n(G,S)$ be the maximum number of winning configurations over all strategies.
For $n\ge 2$ we define the \emph{guessing number}
\[
  \gn(G,S,n):=\log_n W_n(G,S),
  \qquad
  \gn(G,S):=\sup_{n\ge 2}\gn(G,S,n).
\]
When $S=\emptyset$ we abbreviate $W_n(G):=W_n(G,\emptyset)$ and $\gn(G,n):=\gn(G,\emptyset,n)$.
\end{definition}

\begin{remark}[Sources versus self-loops]
If one prefers to stay within the classical definition where every vertex is required to guess its own value,
one may add a self-loop $s\to s$ for each $s\in S$ and then \emph{fix} the local rule at $s$ to be the identity.
This yields the same set of winning configurations.
We use the source-augmented formulation because it avoids suggesting that a player can see its own hat colour.
\end{remark}

\begin{definition}[Labelled guessing game]
\label{def:labelled-guessing-game}
A \emph{labelled digraph} is a tuple $(G,\lambda,S)$ where $G=(V,E)$ is a digraph, $S\subseteq V$ is a set of
source vertices, and $\lambda:V\to X$ is a surjection onto a finite set $X$ (the \emph{labels}).
Vertices in the same fibre $\lambda^{-1}(x)$ are thought of as sharing a common value.

Fix an alphabet $\A$ with $|\A|=n$.
A configuration is an assignment $a:X\to \A$, inducing vertex colours $a(\lambda(v))$.
A strategy assigns to each $v\in V\setminus S$ a local function $g_v:\A^{N^-(v)}\to\A$.
The configuration $a$ is \emph{winning} if for every $v\in V\setminus S$,
\[
  a\bigl(\lambda(v)\bigr)\ =\ g_v\bigl((a(\lambda(u)))_{u\in N^-(v)}\bigr).
\]
We write $W_n(G,\lambda,S)$ for the maximum number of winning configurations, and for $n\ge 2$ set
\[
  \gn(G,\lambda,S,n):=\log_n W_n(G,\lambda,S),
  \qquad
  \gn(G,\lambda,S):=\sup_{n\ge 2}\gn(G,\lambda,S,n).
\]
\end{definition}

In both games $1\le W_n\le n^{|X|}$ (with $X=V$ in the unlabelled case): a configuration is
determined by its $|X|$ values, and the configuration that is constant with value $a_0$ wins
for the strategy whose local functions all return $a_0$.  Hence $0\le\gn(\cdot,n)\le|X|$ and
the suprema above are finite.

\begin{remark}
When $\lambda$ is the identity map and $S=\emptyset$, Definition~\ref{def:labelled-guessing-game} reduces to the
usual guessing game on a digraph as in \cite{riis2006information}.
The labelled variant is an ``identified-value'' extension: several constraint vertices may share the same
underlying label.  This is the natural model for normal-form term systems in which a variable appears as the
output of more than one equation.
\end{remark}

\subsection{Product constructions and convergence of guessing numbers}
\label{subsec:gn-convergence}

The product construction is the basic mechanism behind the existence of a limiting guessing number.
We record it in a form that applies both to the source-augmented game of Definition~\ref{def:guessing-game} and to
its labelled extension (Definition~\ref{def:labelled-guessing-game}).

\begin{proposition}[Monotonicity and product construction]
\label{prop:gn-monotone-product}
Let $(G,S)$ be a digraph with sources as in Definition~\ref{def:guessing-game}, and let $(G,\lambda,S)$ be a labelled
digraph as in Definition~\ref{def:labelled-guessing-game}.
Then:

\smallskip
\noindent\textup{(a) Monotonicity.}
If $m\ge n\ge 1$ then
\[
  W_m(G,S)\ge W_n(G,S)
  \qquad\text{and}\qquad
  W_m(G,\lambda,S)\ge W_n(G,\lambda,S).
\]

\smallskip
\noindent\textup{(b) Product lower bound.}
For all $n_1,n_2\ge 1$,
\[
  W_{n_1n_2}(G,S)\ge W_{n_1}(G,S)\,W_{n_2}(G,S)
  \qquad\text{and}\qquad
  W_{n_1n_2}(G,\lambda,S)\ge W_{n_1}(G,\lambda,S)\,W_{n_2}(G,\lambda,S).
\]
\end{proposition}

\begin{proof}
We prove the labelled statement; the unlabelled case is obtained by taking $\lambda$ to be the identity.

\smallskip
\noindent\emph{(a) Monotonicity.}
Let $\A_n\subseteq \A_m$ with $\abs{\A_n}=n$ and $\abs{\A_m}=m$.
Take a strategy on $\A_n$ attaining $W_n(G,\lambda,S)$ and extend each local function to $\A_m$ by using the same rule
on inputs from $\A_n$ and defining arbitrary outputs otherwise.
Every winning configuration $a:X\to \A_n$ then yields a winning configuration $a:X\to \A_m$ by inclusion, so
$W_m(G,\lambda,S)\ge W_n(G,\lambda,S)$.

\smallskip
\noindent\emph{(b) Product construction.}
Let $\A_1,\A_2$ be alphabets of sizes $n_1,n_2$ and let $g^{(i)}$ be strategies attaining $W_{n_i}(G,\lambda,S)$ on $\A_i$.
On $\A:=\A_1\times \A_2$ define a strategy $g$ by
\[
  g_v\bigl((a_u)_{u\in N^-(v)}\bigr)
  :=
  \Bigl(
    g^{(1)}_v\bigl((\pi_1(a_u))_{u\in N^-(v)}\bigr),\ 
    g^{(2)}_v\bigl((\pi_2(a_u))_{u\in N^-(v)}\bigr)
  \Bigr),
\]
where $\pi_i:\A\to \A_i$ are the coordinate projections.
A configuration $a:X\to \A$ is winning for $g$ if and only if $\pi_1\circ a$ is winning for $g^{(1)}$ and $\pi_2\circ a$
is winning for $g^{(2)}$.
Hence $g$ has $W_{n_1}(G,\lambda,S)\,W_{n_2}(G,\lambda,S)$ winning configurations, proving the inequality.
\end{proof}

\begin{theorem}[Convergence of guessing numbers]
\label{thm:gn-converges}
For every digraph with sources $(G,S)$ and every labelled digraph $(G,\lambda,S)$, the limit
\[
  \lim_{n\to\infty}\gn(\cdot,n)
\]
exists.
More precisely,
\[
  \lim_{n\to\infty}\gn(G,S,n)=\gn(G,S)
  \qquad\text{and}\qquad
  \lim_{n\to\infty}\gn(G,\lambda,S,n)=\gn(G,\lambda,S).
\]
In particular, the suprema in Definitions~\ref{def:guessing-game} and \ref{def:labelled-guessing-game} are in fact
limits.
\end{theorem}

\begin{proof}
We prove the labelled case; the unlabelled case is identical.

Let
\(
  \alpha:=\gn(G,\lambda,S)=\sup_{n\ge 2}\gn(G,\lambda,S,n).
\)
Fix $\varepsilon>0$ and choose $q\ge 2$ with
\(
  \gn(G,\lambda,S,q)\ge \alpha-\varepsilon.
\)
For $n\ge q$, let $t:=\lfloor \log_q n\rfloor$, so $q^t\le n<q^{t+1}$.
By Proposition~\ref{prop:gn-monotone-product}(a) and (b),
\[
  W_n(G,\lambda,S)\ \ge\ W_{q^t}(G,\lambda,S)\ \ge\ W_q(G,\lambda,S)^t.
\]
Therefore
\[
  \gn(G,\lambda,S,n)
  \;=\;
  \log_n W_n(G,\lambda,S)
  \;\ge\;
  \log_n\bigl(W_q(G,\lambda,S)^t\bigr)
  \;=\;
  \frac{t\log q}{\log n}\,\gn(G,\lambda,S,q).
\]
Since $n<q^{t+1}$ we have $\log n<(t+1)\log q$, hence $\frac{t\log q}{\log n}>\frac{t}{t+1}$;
as $\gn(G,\lambda,S,q)\ge 0$, this gives
\begin{equation}\label{eq:gn-floorlog}
  \gn(G,\lambda,S,n)\ \ge\ \frac{t}{t+1}\,\gn(G,\lambda,S,q)
  \ \ge\ \frac{t}{t+1}\,(\alpha-\varepsilon)\qquad(t=\lfloor\log_q n\rfloor,\ n\ge q\ge 2).
\end{equation}
As $n\to\infty$ we have $t\to\infty$, so $\frac{t}{t+1}\to 1$, and therefore
\[
  \liminf_{n\to\infty}\gn(G,\lambda,S,n)\ \ge\ \alpha-\varepsilon.
\]
Since $\varepsilon>0$ was arbitrary, $\liminf\ge \alpha$.
The reverse inequality $\limsup\le \alpha$ is immediate from the definition of $\alpha$.
Hence the limit exists and equals $\alpha$.
\end{proof}

\begin{remark}[Historical note]
For ordinary guessing games (no labels and no source vertices), the limit is known to exist.
Christofides--Markstr\"om~\cite[Theorem~3.6]{christofides2011guessing} prove this for
undirected graphs by the same monotonicity/product argument, and Cameron--Dang--Riis
\cite{cameron2016guessing} state the result for every digraph. The alphabet-extension bounds
of Gadouleau--Riis~\cite[Corollary~3]{gadouleau2011graph}, obtained from guessing graphs,
also imply the same conclusion.
The short product-based argument above applies directly to our source-augmented and labelled variants.
\end{remark}

\subsection{Equivalence with diversified Term Coding}

We now relate diversified normal-form instances to guessing games.
The functional case yields an ordinary guessing game on the dependency digraph; the general case yields a labelled
variant.

\begin{proposition}[Diversified functional instances are guessing games]
\label{prop:div-is-guessing}
Let $\Gamma^{\mathrm{div}}$ be a diversified normal-form instance in functional normal form, and let
$G_{\Gamma}$ be its dependency digraph (Definition~\ref{def:dep-graph}).
Then for every $n\ge 1$,
\[
  \Sn(\Gamma^{\mathrm{div}})\ =\ W_n\bigl(G_{\Gamma},\Src(\Gamma^{\mathrm{div}})\bigr).
\]
\end{proposition}

\begin{proof}
Fix an alphabet $\A$ with $|\A|=n$ and write $S:=\Src(\Gamma^{\mathrm{div}})$.

\smallskip
\noindent\emph{From interpretations to strategies.}
Let $\mathcal I$ be an interpretation of the diversified signature on $\A$.
Since $\Gamma^{\mathrm{div}}$ is in functional normal form, every $x_j\notin S$ has a unique defining equation,
which is either of the form
\(
  f^{(\ell)}(x_{i_1},\dots,x_{i_r})=x_j
\)
or of the form \(c^{(\ell)}=x_j\).
Define a strategy on $(G_{\Gamma},S)$ by setting, for each $x_j\notin S$,
\[
  g_{x_j}\bigl((a_u)_{u\in N^-(x_j)}\bigr)
  := f^{(\ell)\,\mathcal I}(a_{i_1},\dots,a_{i_r})
\]
in the first case, with repeated and ordered arguments read through the argument-position map of
Remark~\ref{rem:argument-positions}, and by taking $g_{x_j}$ to be the constant map with value
$(c^{(\ell)})^{\mathcal I}$ in the second.
Then an assignment $\mathbf a\in\A^V$ is a solution of $\Gamma^{\mathrm{div}}$ under $\mathcal I$ if and only if it
is a winning configuration for this strategy.
Hence
\(
  \abs{S_{\mathcal I}(\Gamma^{\mathrm{div}})}\le W_n(G_{\Gamma},S)
\),
and taking the maximum over $\mathcal I$ gives
\(
  \Sn(\Gamma^{\mathrm{div}})\le W_n(G_{\Gamma},S).
\)

\smallskip
\noindent\emph{From strategies to interpretations.}
Conversely, let $g$ be a strategy on $(G_{\Gamma},S)$.
For each non-constant equation
\(
  f^{(\ell)}(x_{i_1},\dots,x_{i_r})=x_j
\)
define the interpretation of the symbol $f^{(\ell)}$ on consistent argument tuples by the corresponding local
function $g_{x_j}$ and extend it arbitrarily to all of $\A^r$, as in
Remark~\ref{rem:argument-positions}.
This is consistent because $\Gamma^{\mathrm{div}}$ is diversified: each symbol $f^{(\ell)}$ appears in exactly one
equation.

For a diversified constant equation $c^{(\ell)}=x_j$, interpret the fresh constant $c^{(\ell)}$ as the value of
the nullary local rule at $x_j$.
With this interpretation $\mathcal I$, the winning configurations of $g$ are exactly the solutions of
$\Gamma^{\mathrm{div}}$ under $\mathcal I$.
Thus $W_n(G_{\Gamma},S)\le \Sn(\Gamma^{\mathrm{div}})$, proving equality.
\end{proof}

\begin{definition}[Labelled dependency graph]
\label{def:labelled-dep-graph}
Let $\Gamma^{\mathrm{div}}$ be a diversified normal-form instance on variable set
$V=\{x_1,\dots,x_k\}$.
We define its \emph{labelled dependency graph} $\widetilde G_{\Gamma}$ to be the labelled digraph
$(\widetilde G_{\Gamma},\lambda_{\Gamma},S_{\Gamma})$ given as follows.

\begin{itemize}
\item The vertex set is a disjoint union $\widetilde V=V^{\mathrm{var}}\sqcup V^{\mathrm{eq}}$, where
$V^{\mathrm{var}}:=\{v_1,\dots,v_k\}$ has one \emph{variable vertex} $v_i$ for each $x_i$, and $V^{\mathrm{eq}}$ has one
\emph{equation vertex} $e$ for each equation of $\Gamma^{\mathrm{div}}$.
\item The label map $\lambda_{\Gamma}:\widetilde V\to V$ sends $v_i\mapsto x_i$, and sends an equation vertex
corresponding to an equation with right-hand side $x_j$ to the label $x_j$.
\item The source set is $S_{\Gamma}:=V^{\mathrm{var}}$ (variable vertices impose no guessing condition).
\item For an equation vertex corresponding to
$f^{(\ell)}(x_{i_1},\dots,x_{i_r})=x_j$ we add directed edges $v_{i_p}\to e$ for $p=1,\dots,r$.
For a diversified constant equation $c^{(\ell)}=x_j$ the corresponding equation vertex has no in-neighbours.
\end{itemize}
\end{definition}

\begin{proposition}[Diversified instances are labelled guessing games]
\label{prop:div-is-labelled-guessing}
Let $\Gamma^{\mathrm{div}}$ be a diversified normal-form instance and let $\widetilde G_{\Gamma}$ be its labelled
dependency graph (Definition~\ref{def:labelled-dep-graph}).
Then for every $n\ge 1$,
\[
  \Sn(\Gamma^{\mathrm{div}})\ =\ W_n\bigl(\widetilde G_{\Gamma},\lambda_{\Gamma},S_{\Gamma}\bigr).
\]
\end{proposition}

\begin{proof}
Fix an alphabet $\A$ with $|\A|=n$.
A configuration in the labelled guessing game on $\widetilde G_{\Gamma}$ is exactly an assignment
$a:V\to\A$ of values to the term variables.

Given an interpretation $\mathcal I$ of the diversified signature on $\A$, define a strategy on
$(\widetilde G_{\Gamma},\lambda_{\Gamma},S_{\Gamma})$ by assigning to each equation vertex the local function given by
the interpretation of the symbol occurring in that equation, precomposed with the argument-position map as in
Remark~\ref{rem:argument-positions}.
(For a diversified constant equation $c^{(\ell)}=x_j$, this is the constant map with value
$(c^{(\ell)})^{\mathcal I}$.)
Then $a$ is winning if and only if it satisfies every equation of $\Gamma^{\mathrm{div}}$ under $\mathcal I$.
Hence \(\abs{S_{\mathcal I}(\Gamma^{\mathrm{div}})}\le W_n(\widetilde G_{\Gamma},\lambda_{\Gamma},S_{\Gamma})\), and
maximising over $\mathcal I$ gives
\(\Sn(\Gamma^{\mathrm{div}})\le W_n(\widetilde G_{\Gamma},\lambda_{\Gamma},S_{\Gamma})\).

Conversely, given a strategy on $(\widetilde G_{\Gamma},\lambda_{\Gamma},S_{\Gamma})$, interpret each diversified
non-constant symbol $f^{(\ell)}$ by extending the local function at the corresponding equation vertex from
consistent argument tuples to all tuples, and interpret each fresh diversified constant by its nullary local rule.
With this interpretation, winning configurations are exactly solutions of $\Gamma^{\mathrm{div}}$.
Thus the two maxima coincide.
\end{proof}

\begin{remark*}
Conversely, every finite labelled digraph $(G,\lambda,S)$ arises in this way: give each
non-source vertex $v$ one equation $f_v(\lambda(u_1),\dots,\lambda(u_r))=\lambda(v)$, where
$(u_1,\dots,u_r)$ enumerates $N^-(v)$ and $f_v$ is a fresh symbol.  The winning configurations
of the labelled game on $(G,\lambda,S)$ are exactly the solutions of this diversified system,
so Proposition~\ref{prop:div-is-labelled-guessing} identifies diversified normal-form instances
with labelled guessing games, and Proposition~\ref{prop:gn-monotone-product} and
Theorem~\ref{thm:gn-converges} are statements about one and the same class of objects.
\end{remark*}

\begin{proposition}[Unlabelled upper bound for non-functional instances]
\label{prop:div-upper}
Let $\Gamma^{\mathrm{div}}$ be a diversified normal-form instance with dependency digraph $G_{\Gamma}$.
Then for every $n\ge 1$,
\[
  \Sn(\Gamma^{\mathrm{div}})\ \le\ W_n\bigl(G_{\Gamma},\Src(\Gamma^{\mathrm{div}})\bigr).
\]
\end{proposition}

\begin{proof}
Fix $n$ and an interpretation $\mathcal I$ of the diversified signature on an alphabet $\A$.
For each non-source variable $x_j\notin\Src(\Gamma^{\mathrm{div}})$, choose \emph{one} equation of
$\Gamma^{\mathrm{div}}$ with right-hand side $x_j$, say
\(f^{(\ell)}(x_{i_1},\dots,x_{i_r})=x_j\) (or $c^{(\ell)}=x_j$).
Define a strategy on $(G_{\Gamma},\Src(\Gamma^{\mathrm{div}}))$ by taking $g_{x_j}$ to be the corresponding
interpretation of $f^{(\ell)}$, precomposed with projection from the full in-neighbour set to the ordered argument
positions of the chosen equation (or the constant map with value $(c^{(\ell)})^{\mathcal I}$).
Every solution of $\Gamma^{\mathrm{div}}$ under $\mathcal I$ satisfies, in particular, these chosen equations, and is
therefore a winning configuration for this strategy.
Hence
\(
  \abs{S_{\mathcal I}(\Gamma^{\mathrm{div}})}\le W_n(G_{\Gamma},\Src(\Gamma^{\mathrm{div}})).
\)
Taking the maximum over $\mathcal I$ proves the claim.
\end{proof}

\begin{remark*}
The inequality can be strict.  For the diversified system $f_1(u)=z$, $f_2(v)=z$, $f_3(z)=u$
with source $v$, a winning configuration of the labelled game is determined by $v$ (then
$z=g_2(v)$ and $u=g_3(z)$ are forced and must satisfy $g_1(u)=z$), and $n$ winning
configurations are attained by identity rules, so $\Sn(\Gamma^{\mathrm{div}})=n$; in the
ordinary game on the digraph with edges
$u\to z$, $v\to z$, $z\to u$ and source $v$, the single rule $g_z(u,v)=u$ together with
$g_u(z)=z$ wins on all $n^2$ configurations with $u=z$.
\end{remark*}

\section{The guessing-number sandwich theorem}
\label{sec:sandwich}

We now combine the reduction pipeline (Proposition~\ref{prop:normalisation}), the diversification sandwich
(Lemma~\ref{lem:div-sandwich}), and the guessing-game equivalence results of Section~\ref{sec:guessing}.

We state the functional case first; the labelled theorem, which applies to every collision-free
normal form, follows.

\begin{proposition}[Guessing-number sandwich theorem (functional case)]
\label{prop:sandwich-unlabelled}
Let $\Gamma$ be a Term Coding instance and let $\Gamma^{\mathrm{nf}}$ be the normal-form instance produced by
Proposition~\ref{prop:normalisation}.
In particular, $\Gamma^{\mathrm{nf}}$ satisfies the Term-DAG/no-collision property of Definition~\ref{def:term-dag}.
Assume that $\Gamma^{\mathrm{nf}}$ is in functional normal form (Definition~\ref{def:fnf}), and let
$\Gamma^{\mathrm{div}}$ be its diversification.
Let $k$ be the number of variables in $\Gamma^{\mathrm{nf}}$, let $G_\Gamma$ be the dependency digraph
(Definition~\ref{def:dep-graph}), and put $S:=\Src(\Gamma^{\mathrm{nf}})$.
Then for every $n\ge k$,
\[
  W_{\lfloor n/k\rfloor}(G_\Gamma,S)
  \ \le\ \Sn(\Gamma)
  \ \le\ W_n(G_\Gamma,S).
\]
In particular, for every $n\ge 2k$ (so that $\lfloor n/k\rfloor\ge 2$),
\[
  \lfloor n/k\rfloor^{\gn(G_\Gamma,S,\lfloor n/k\rfloor)}
  \ \le\ \Sn(\Gamma)
  \ \le\ n^{\gn(G_\Gamma,S,n)}.
\]
\end{proposition}

\begin{proof}
Normalisation and Lemma~\ref{lem:div-sandwich} give, for $n\ge k$ and
$m:=\lfloor n/k\rfloor$,
\[
  S_m(\Gamma^{\mathrm{div}})\ \le\ \Sn(\Gamma^{\mathrm{nf}})
  \ \le\ \Sn(\Gamma^{\mathrm{div}}).
\]
In functional normal form, Proposition~\ref{prop:div-is-guessing} identifies the two
diversified endpoints with $W_m(G_\Gamma,S)$ and $W_n(G_\Gamma,S)$. The final display is
the definition of $\gn$.
\end{proof}

\begin{theorem}[Generalised guessing-number sandwich theorem (labelled form)]
\label{thm:sandwich-labelled}
Let $\Gamma$ be a Term Coding instance and let $\Gamma^{\mathrm{nf}}$ be the normal-form instance produced by
Proposition~\ref{prop:normalisation} on variable set $V=\{x_1,\dots,x_k\}$.
Let $\Gamma^{\mathrm{div}}$ be its diversification, and let
$(\widetilde G_{\Gamma},\lambda_{\Gamma},S_{\Gamma})$ be the labelled dependency graph of
$\Gamma^{\mathrm{div}}$ (Definition~\ref{def:labelled-dep-graph}).
Then for every $n\ge k$,
\[
  W_{\lfloor n/k\rfloor}(\widetilde G_{\Gamma},\lambda_{\Gamma},S_{\Gamma})
  \ \le\ \Sn(\Gamma)
  \ \le\ W_n(\widetilde G_{\Gamma},\lambda_{\Gamma},S_{\Gamma}).
\]
\end{theorem}

\begin{proof}
Proposition~\ref{prop:normalisation} lets us work with $\Gamma^{\mathrm{nf}}$.
For $n\ge k$ and $m:=\lfloor n/k\rfloor$, Lemma~\ref{lem:div-sandwich} gives
\[
  S_m(\Gamma^{\mathrm{div}})\ \le\ \Sn(\Gamma^{\mathrm{nf}})\ \le\ \Sn(\Gamma^{\mathrm{div}}).
\]
Proposition~\ref{prop:div-is-labelled-guessing} identifies
$S_t(\Gamma^{\mathrm{div}})=W_t(\widetilde G_{\Gamma},\lambda_{\Gamma},S_{\Gamma})$ for every $t$.
Substituting $t=m$ and $t=n$ yields the claimed sandwich.
\end{proof}

\begin{corollary}[Exponent and $o(1)$-tight asymptotics]
\label{cor:exponent}
Let $\Gamma$ be a Term Coding instance and let $\Gamma^{\mathrm{nf}}$ be the collision-free normal form of
Proposition~\ref{prop:normalisation} on $k$ variables.
Let $\Gamma^{\mathrm{div}}$ be its diversification and let
$(\widetilde G_{\Gamma},\lambda_{\Gamma},S_{\Gamma})$ be the labelled dependency graph of
$\Gamma^{\mathrm{div}}$ (Definition~\ref{def:labelled-dep-graph}).

Define
\[
  \alpha(\Gamma):=\gn\bigl(\widetilde G_{\Gamma},\lambda_{\Gamma},S_{\Gamma}\bigr).
\]
If $\Gamma^{\mathrm{nf}}$ is in functional normal form, then equivalently
\[
  \alpha(\Gamma)=\gn\bigl(G_\Gamma,\Src(\Gamma^{\mathrm{nf}})\bigr),
\]
where $G_\Gamma$ is the (unlabelled) dependency digraph.

Then, as $n\to\infty$,
\[
  \logn \Sn(\Gamma)=\alpha(\Gamma)+o(1),
\]
or equivalently $\Sn(\Gamma)=n^{\alpha(\Gamma)+o(1)}$.
In particular, the limit $\lim_{n\to\infty}\logn\Sn(\Gamma)$ exists and equals $\alpha(\Gamma)$.
\end{corollary}

\begin{proof}
Let $k$ be the number of variables in $\Gamma^{\mathrm{nf}}$.
For every $n\ge k$, Theorem~\ref{thm:sandwich-labelled} yields
\[
  W_{\lfloor n/k\rfloor}(\widetilde G_{\Gamma},\lambda_{\Gamma},S_{\Gamma})
  \ \le\ \Sn(\Gamma)
  \ \le\ W_n(\widetilde G_{\Gamma},\lambda_{\Gamma},S_{\Gamma}).
\]
Taking $\log_n$ gives
\[
  \frac{\log \lfloor n/k\rfloor}{\log n}\,\gn\bigl(\widetilde G_{\Gamma},\lambda_{\Gamma},S_{\Gamma},\lfloor n/k\rfloor\bigr)
  \ \le\ 
  \logn \Sn(\Gamma)
  \ \le\
  \gn\bigl(\widetilde G_{\Gamma},\lambda_{\Gamma},S_{\Gamma},n\bigr).
\]
As $n\to\infty$ we have $\frac{\log \lfloor n/k\rfloor}{\log n}\to 1$, and by
Theorem~\ref{thm:gn-converges} we have
\(
  \gn(\widetilde G_{\Gamma},\lambda_{\Gamma},S_{\Gamma},n)\to
  \gn(\widetilde G_{\Gamma},\lambda_{\Gamma},S_{\Gamma})=\alpha(\Gamma)
\)
and also
\(
  \gn(\widetilde G_{\Gamma},\lambda_{\Gamma},S_{\Gamma},\lfloor n/k\rfloor)\to \alpha(\Gamma).
\)
The squeeze theorem now gives $\logn \Sn(\Gamma)\to \alpha(\Gamma)$, i.e.\ $\logn\Sn(\Gamma)=\alpha(\Gamma)+o(1)$.

If $\Gamma^{\mathrm{nf}}$ is in functional normal form, the labelled game collapses to the ordinary guessing game on
$G_\Gamma$ with sources $\Src(\Gamma^{\mathrm{nf}})$ by Proposition~\ref{prop:div-is-guessing}.
\end{proof}

\section{Entropy and polymatroid viewpoint}
\label{sec:entropy}

The sandwich theorem is combinatorial, but it is compatible with the entropy inequalities used to bound guessing
numbers.
This section records a compact ``entropy dictionary'' for term coding.

\subsection{Uniform distribution on the solution set}

Fix an interpretation $\mathcal I$ of $\Gamma$ over an alphabet $\A$ of size $n$.
Let $X=(X_1,\dots,X_v)$ be a random assignment drawn uniformly from the solution set
$\SI(\Gamma)\subseteq\A^v$.
Then the joint entropy satisfies
\begin{equation}
\label{eq:entropy-logsize}
  H(X_1,\dots,X_v)=\log \abs{\SI(\Gamma)}
\end{equation}
(with base-$e$ logarithms).
Normalising by $\log n$ gives
\[
  \frac{1}{\log n}H(X_1,\dots,X_v)=\logn\abs{\SI(\Gamma)}.
\]
Equivalently, if we measure entropy in base-$n$ units by $H_n(\cdot):=H(\cdot)/\log n$, then
$H_n(X_1,\dots,X_v)=\log_n\abs{\SI(\Gamma)}$.
Thus maximising code size is equivalent to maximising the normalised joint entropy of a uniform-on-code
random vector.

\subsection{Term equations as functional dependencies}

Each normal-form equation $f(\mathbf U)=V$ in $\Gamma^{\mathrm{nf}}$ forces a deterministic dependence:
$V$ is a function of $\mathbf U$ on the solution set.
Equivalently,
\[
  H(V\mid \mathbf U)=0.
\]
Therefore, any upper bound on the joint entropy that follows from Shannon-type inequalities,
the linear constraints $H(V\mid\mathbf U)=0$, and the alphabet bounds $H(X_S)\le|S|\log n$
yields an upper bound on $\logn\abs{\SI(\Gamma)}$.

\subsection{Polymatroids and the Shannon cone}

For a random vector $(X_1,\dots,X_v)$ define the set function $h(S)=H(X_S)$.
Shannon's inequalities imply that $h$ is a \emph{polymatroid rank function}:
\begin{enumerate}[leftmargin=2.2em]
\item $h(S)\ge 0$ and $h(\emptyset)=0$,
\item $h$ is monotone: $h(S)\le h(T)$ for $S\subseteq T$,
\item $h$ is submodular: $h(S)+h(T)\ge h(S\cap T)+h(S\cup T)$.
\end{enumerate}
The set of all such $h$ forms the classical Shannon cone $\Gamma_v$.
The almost-entropic cone $\overline{\Gamma_v^*}$ (closure of entropy vectors) is typically smaller,
and non-Shannon inequalities can be required for tight bounds
\cite{zhang1997non,baber2016graph}.

In term coding, the constraints $H(V\mid\mathbf U)=0$ define linear sections of entropy space.
Every code yields an entropic point in the corresponding section, so maximising over the Shannon cone or the
almost-entropic cone gives systematic upper bounds.  Exact identification with a cone optimisation requires a
separate achievability argument; we do not assume such an identification in the sandwich theorem.

\begin{proposition}[The five-cycle entropy polytope has a non-integer vertex]
\label{prop:c5-vertex}
Let $P_5$ be the set of functions $h:2^{\{1,\dots,5\}}\to\mathbb R$ that are polymatroid rank
functions ($h(\emptyset)=0$, monotone, submodular) and satisfy the five-cycle dependence
constraints $h(\{i-1,i,i+1\})=h(\{i-1,i+1\})$ and the alphabet bounds $h(\{i\})\le 1$ for all
$i\in\mathbb Z/5\mathbb Z$.
Then
\[
  \max_{h\in P_5} h(\{1,\dots,5\})\;=\;\frac52,
\]
and the maximum is attained at exactly one point of $P_5$, namely
\[
  h^*(S)\;:=\;\tfrac12\,\bigl|S\cup(S+1)\bigr|\qquad(S\subseteq\mathbb Z/5\mathbb Z),
\]
which is therefore a vertex of the polytope $P_5$.  It has the non-integer values
$h^*(\{i,i+1\})=3/2$ and $h^*(\{1,\dots,5\})=5/2$.
Moreover $h^*$ is entropic: for $n=m^2$ it is the base-$n$ entropy vector of the uniform
distribution on the $m^5$ winning configurations $X_i=(U_i,U_{i+1})$ of the square-alphabet
strategy in the proof of Theorem~\ref{thm:c5}.
\end{proposition}

\begin{proof}
First, $h^*\in P_5$: $S\mapsto|S\cup(S+1)|$ is a coverage function, hence monotone and
submodular; $\{i-1,i,i+1\}$ and $\{i-1,i+1\}$ have the same closure $\{i-1,i,i+1,i+2\}$; and
$|\{i,i+1\}|=2$.  Clearly $h^*(\{1,\dots,5\})=5/2$.

\smallskip
\noindent\emph{Upper bound.}
Let $h\in P_5$ and abbreviate $h(\{1,2,4\})=:h(124)$, etc.
Submodularity for $\{1,2,3,4\}$,\allowbreak\ $\{1,4,5\}$ together with $h(145)=h(14)$ gives
$h(12345)\le h(1234)$; for $\{1,2,4\},\{2,3,4\}$ with $h(234)=h(24)$ it gives
$h(1234)\le h(124)$; and for $\{1,2\},\{4\}$ (using $h(\emptyset)=0$) it gives
$h(124)\le h(12)+h(4)$.  Hence
\begin{equation}\label{eq:c5-upper-1}
  h(12345)\;\le\;h(12)+h(4).
\end{equation}
Similarly, submodularity for $\{1,2,3,5\},\{3,4,5\}$ with $h(345)=h(35)$ gives
$h(12345)\le h(1235)$, and for $\{1,2,3\},\{1,2,5\}$ with $h(123)=h(13)$ and $h(125)=h(25)$ it
gives $h(12)+h(1235)\le h(13)+h(25)$; using submodularity for the disjoint pairs
$\{1\},\{3\}$ and $\{2\},\{5\}$ and the alphabet bounds,
\begin{equation}\label{eq:c5-upper-2}
  h(12)+h(12345)\;\le\;h(13)+h(25)\;\le\;h(1)+h(3)+h(2)+h(5)\;\le\;4.
\end{equation}
Adding \eqref{eq:c5-upper-1} and \eqref{eq:c5-upper-2} and using $h(4)\le1$ yields
$2h(12345)\le 5$.

\smallskip
\noindent\emph{Uniqueness.}
If $h(12345)=5/2$, every inequality above is tight: \eqref{eq:c5-upper-1} and
\eqref{eq:c5-upper-2} force $h(12)=3/2$, $h(1)=\dots=h(5)=1$, $h(13)=h(25)=2$ and
$h(124)=h(1234)=h(1235)=5/2$, and the dependence identities give $h(123)=h(125)=2$.
Applying the same argument to the four rotations $i\mapsto i+r$ of the chain pins every
coordinate to $h^*(S)$.  Thus $h^*$ is the unique maximiser of a linear functional over the
convex set $P_5$, hence an extreme point, i.e.\ a vertex.

\smallskip
\noindent\emph{Entropic realisation.}
Let $U=(U_1,\dots,U_5)$ be uniform on $[m]^5$ and $X_i:=(U_i,U_{i+1})$.  For the strategy in
the proof of Theorem~\ref{thm:c5}, a configuration $\bigl((U_i,V_i)\bigr)_i$ is winning if and
only if $V_i=U_{i+1}$ for all $i$ (this already forces $U_i=V_{i-1}$), so $U\mapsto X$ is a
bijection onto the winning set, which has exactly $m^5$ elements, and $X$ is uniform on it.
For $S\subseteq\mathbb Z/5\mathbb Z$ the projection $X_S$ is a bijective function of
$(U_j)_{j\in S\cup(S+1)}$, which is uniform on $[m]^{S\cup(S+1)}$; hence
$H(X_S)=|S\cup(S+1)|\log m=h^*(S)\log n$.
\end{proof}

\begin{remark}[Why non-integer exponents appear]
A matroid rank function takes integer values; polymatroids and entropy vectors need not.
Proposition~\ref{prop:c5-vertex} shows that for the five-cycle the entropy optimum is attained
at a genuinely non-integer vertex of the constraint polytope, and Theorem~\ref{thm:c5} shows
that this vertex governs the exponent $5/2$.
\end{remark}

\section{Applications and case studies}
\label{sec:applications}

A central purpose of this article is to show that the framework yields concrete information on familiar
structures in extremal combinatorics.
We therefore record three illustrative case studies.

\subsection{The five-cycle and a non-integer exponent}
\label{subsec:c5-exponent}

Let $C_5$ denote the bidirected cycle on $5$ vertices.
Section~\ref{subsec:c5-visual} exhibited a term-coding instance whose dependency graph is $C_5$.
The key fact is:

\begin{theorem}[$\gn(C_5)=5/2$]
\label{thm:c5}
For the bidirected $5$-cycle $C_5$,
\[\gn(C_5)=\sup_{n\ge 2}\gn(C_5,n)=\frac{5}{2}.
\]
Consequently, the $C_5$ term-coding instance of \eqref{eq:c5-nested} has exponent
$\alpha(\Gamma)=5/2$.
\end{theorem}

\begin{proof}
The value $\gn(C_5)=5/2$ is classical; see \cite[Section~5]{riis2007graphentropy} for an
explicit computation and \cite{christofides2011guessing,gadouleau2011graph} for further discussion.
The upper bound is a consequence of Proposition~\ref{prop:c5-vertex}; we include the explicit
construction for the matching lower bound.

\smallskip
\noindent\emph{Upper bound.}
Fix $n\ge 2$ and a strategy on $C_5$ over an alphabet $\A$ of size $n$ with winning set
$W\subseteq\A^5$, and let $(X_1,\dots,X_5)$ be uniform on $W$.
Put $h(S):=H(X_S)/\log n$ for $S\subseteq\{1,\dots,5\}$.
Then $h(\emptyset)=0$ and $h$ is monotone and submodular (Section~\ref{sec:entropy}); each
local winning condition makes $X_i$ a function of $(X_{i-1},X_{i+1})$, so
$h(\{i-1,i,i+1\})=h(\{i-1,i+1\})$; and $h(\{i\})\le 1$ because $X_i$ takes at most $n$ values.
Thus $h\in P_5$, and Proposition~\ref{prop:c5-vertex} gives
$\log|W|=H(X_1,\dots,X_5)\le\tfrac52\log n$, i.e.\ $\gn(C_5,n)\le 5/2$ for all $n$.

\smallskip
\noindent\emph{Lower bound.}
When $n=m^2$, identify $\A$ with $\{1,\dots,m\}^2$ and write $X_i=(U_i,V_i)$.
Define local update rules by
\[
  \widehat X_i:=\bigl(V_{i-1},\,U_{i+1}\bigr)\qquad (i\in \mathbb Z/5\mathbb Z).
\]
A configuration $\bigl((U_i,V_i)\bigr)_i$ is winning if and only if $V_i=U_{i+1}$ for all $i$
(the condition $U_i=V_{i-1}$ is then automatic), so the winning configurations are exactly the
$m^5=n^{5/2}$ tuples $\bigl((U_i,U_{i+1})\bigr)_i$ with $(U_1,\dots,U_5)\in[m]^5$.
Hence $W_n(C_5)\ge n^{5/2}$ for square $n$, so $\gn(C_5)\ge 5/2$.
Together with the upper bound this yields $\gn(C_5)=5/2$.
Finally, the normal form \eqref{eq:c5-nf} of \eqref{eq:c5-nested} is in functional normal form,
has no source variables, and has dependency digraph $C_5$ (Section~\ref{subsec:c5-visual}), so
Corollary~\ref{cor:exponent} together with Proposition~\ref{prop:sandwich-unlabelled} gives
$\alpha(\Gamma)=\gn(C_5)=5/2$.
\end{proof}

\begin{remark}
This example is the simplest place where ``dimension'' is non-integer.
In linear settings, solution spaces often behave like $n^d$ for integer $d$.
Here the exponent $5/2$ is forced by entropy and is not a matroid rank.
\end{remark}

\subsection{Self-orthogonal Latin squares and formulation sensitivity}
\label{subsec:sols}

A \emph{self-orthogonal Latin square} (SOLS) of order $n$ is an $n\times n$ array over $\A$ such that every row
and column is a permutation (Latin property) and the pair $(f(x,y),f(y,x))$ uniquely identifies $(x,y)$
(self-orthogonality).
SOLS are known to exist for all $n\notin\{2,3,6\}$ \cite{brayton1974self,hedayat1975}.

\medskip
\noindent
This case study illustrates two distinct (but related) messages.
First, the ``standard'' SOLS identities give an exact term-coding formulation of the existence problem.
Second, a natural \emph{self-decoding} strengthening is universally inconsistent, yet still yields a
nontrivial optimisation problem whose \emph{numerics depend on formulation choices} (which variables are treated
as free inputs).

\paragraph{Standard SOLS as a term-coding instance.}
Let $f,h_1,h_2,h_3,h_4:\A^2\to\A$ and impose the equations
\begin{equation}
\label{eq:sols}
\begin{aligned}
  &h_1\bigl(f(x,y),y\bigr)=x,\qquad
  &&h_2\bigl(x,f(x,y)\bigr)=y,\\
  &h_3\bigl(f(x,y),f(y,x)\bigr)=x,\qquad
  &&h_4\bigl(f(x,y),f(y,x)\bigr)=y.
\end{aligned}
\end{equation}
Then $\Sn(\Gamma_{\mathrm{SOLS}})=n^2$ if and only if a SOLS of order $n$ exists.
(Indeed, the first two identities encode the Latin property via two decoding maps, and the last two encode
self-orthogonality via decoding from the ordered pair $(f(x,y),f(y,x))$.)

\begin{lemma}[Partial SOLS viewpoint]
\label{lem:sols-partial}
Fix an alphabet $\A$ and an interpretation $\mathcal I$ of the five symbols in \eqref{eq:sols}.
Let
\[
  C_{\mathcal I}:=\bigl\{(x,y)\in \A^2:\ \eqref{eq:sols}\ \text{holds at }(x,y)\bigr\}.
\]
Then $f$ behaves like a \emph{partial Latin square} on $C_{\mathcal I}$ and the superposition map is injective
on $C_{\mathcal I}$:
\begin{enumerate}[label=(\roman*),leftmargin=2.2em]
\item for each fixed $y\in\A$, the map $x\mapsto f(x,y)$ is injective on $\{x:(x,y)\in C_{\mathcal I}\}$;
\item for each fixed $x\in\A$, the map $y\mapsto f(x,y)$ is injective on $\{y:(x,y)\in C_{\mathcal I}\}$;
\item the map $\phi:C_{\mathcal I}\to \A^2$ defined by $\phi(x,y)=(f(x,y),f(y,x))$ is injective.
\end{enumerate}
Conversely, if $f:\A^2\to\A$ and $C\subseteq\A^2$ satisfy (i)--(iii) (with $C$ in place of $C_{\mathcal I}$),
then there exist maps $h_1,\dots,h_4:\A^2\to\A$ such that every pair in $C$ satisfies \eqref{eq:sols}.
In particular, $\Sn(\Gamma_{\mathrm{SOLS}})$ is the maximum size of a subset $C\subseteq\A^2$ for which such
an $f$ exists.
\end{lemma}

\begin{proof}
Let $(x,y),(x',y)\in C_{\mathcal I}$ and suppose $f(x,y)=f(x',y)$. Applying the first identity of
\eqref{eq:sols} gives
\[
  x=h_1\bigl(f(x,y),y\bigr)=h_1\bigl(f(x',y),y\bigr)=x',
\]
so $x\mapsto f(x,y)$ is injective on the fibre over $y$. This proves (i). The proof of (ii) is identical using
$h_2\bigl(x,f(x,y)\bigr)=y$.
For (iii), if $(x,y),(x',y')\in C_{\mathcal I}$ satisfy
$(f(x,y),f(y,x))=(f(x',y'),f(y',x'))$, then applying the last two identities of \eqref{eq:sols} yields
\[
  x=h_3\bigl(f(x,y),f(y,x)\bigr)=h_3\bigl(f(x',y'),f(y',x')\bigr)=x'
\]
and similarly $y=y'$, so $\phi$ is injective.

Conversely, given $f$ and $C$ with properties (i)--(iii), define $h_1$ and $h_2$ on the relevant inputs by
setting $h_1\bigl(f(x,y),y\bigr):=x$ and $h_2\bigl(x,f(x,y)\bigr):=y$ for each $(x,y)\in C$.
By (i) and (ii) these prescriptions are consistent.
Similarly, define $h_3\bigl(f(x,y),f(y,x)\bigr):=x$ and $h_4\bigl(f(x,y),f(y,x)\bigr):=y$ for each $(x,y)\in C$;
by (iii) this is consistent.
Extend $h_1,\dots,h_4$ arbitrarily to all of $\A^2$.
Then every $(x,y)\in C$ satisfies \eqref{eq:sols} by construction.
\end{proof}

\begin{lemma}[An $r$-SOLS lower bound]
\label{lem:sols-rsols}
Assume $f:\A^2\to\A$ is a Latin square operation (a quasigroup).
Let
\[
  r\ :=\ \abs{\bigl\{(f(x,y),f(y,x)):\ x,y\in\A\bigr\}}.
\]
Then there exist decoding maps $h_1,\dots,h_4:\A^2\to\A$ such that the set of pairs satisfying \eqref{eq:sols}
has size exactly $r$.
Equivalently, if an $r$-self-orthogonal Latin square of order $n$ exists, then
\(\Sn(\Gamma_{\mathrm{SOLS}})\ge r\).
\end{lemma}

\begin{proof}
Since $f$ is a quasigroup, there exist division operations $h_1,h_2$ such that
$h_1\bigl(f(x,y),y\bigr)=x$ and $h_2\bigl(x,f(x,y)\bigr)=y$ hold for \emph{all} $x,y\in\A$.
Let
\(P:=\{(f(x,y),f(y,x)):\ x,y\in\A\}\), so $\abs{P}=r$.
Choose a section $\sigma:P\to\A^2$ (possible since $P$ is finite) such that for each $(u,v)\in P$,
writing $\sigma(u,v)=(x_{u,v},y_{u,v})$ gives
\((f(x_{u,v},y_{u,v}),f(y_{u,v},x_{u,v}))=(u,v)\).
Define $h_3(u,v):=x_{u,v}$ and $h_4(u,v):=y_{u,v}$ for $(u,v)\in P$, and extend $h_3,h_4$ arbitrarily to all of
$\A^2$.
Then each representative pair $(x_{u,v},y_{u,v})$ satisfies the last two identities of \eqref{eq:sols}, and any
other pair $(x,y)$ with the same image $(u,v)$ fails at least one of those identities.
Thus exactly the $r$ chosen representatives satisfy \eqref{eq:sols}.
\end{proof}

\paragraph{A citation-backed bound at the exceptional order $6$.}
While the binary existence question is settled, Term Coding quantifies the ``distance to satisfiability'' at the
impossible orders.  The standard formulation permits embeddings and other partial constructions.
For $n=6$ a perfect score of $36$ is impossible since no SOLS of order $6$ exists
\cite{brayton1974self,hedayat1975}, so \(\Sn(\Gamma_{\mathrm{SOLS}})\le 35\).
The published spectrum of $r$-self-orthogonal Latin squares implies that an $r$-SOLS of order $6$ exists
with $r=31$ (see, e.g., \cite{xu2004spectrum,xu2006existence} and the concise summary
\cite[Table~1(b)]{bereg2024computing}).
By Lemma~\ref{lem:sols-rsols} this gives the lower bound \(\Sn(\Gamma_{\mathrm{SOLS}})\ge 31\).
Thus
\[
  31\ \le\ \Sn(\Gamma_{\mathrm{SOLS}})\ \le\ 35\qquad(n=6).
\]
We do not claim the exact optimum.

\paragraph{A self-decoding strengthening (SDOS).}
A natural (but too strong) variant asks $f$ itself to perform the decoding.
We refer to this as the \emph{self-decoding orthogonal square} (SDOS) variant:

\begin{lemma}
\label{lem:sols-strong-unsat}
There is no interpretation $f:\A^2\to\A$ with $\abs{\A}>1$ satisfying
\begin{equation}
\label{eq:sols-strong}
\begin{aligned}
& f\bigl(f(x,y),y\bigr)=x, \qquad
&& f\bigl(x,f(y,x)\bigr)=y,\\[-2pt]
& f\bigl(f(x,y),f(y,x)\bigr)=x,\qquad
&& f\bigl(f(y,x),f(x,y)\bigr)=y.
\end{aligned}
\end{equation}
\end{lemma}

\begin{proof}
The first two identities force commutativity.  Indeed, the first identity gives
$f(f(z,x),x)=z$, and the second identity, applied with $y=f(z,x)$, gives
\[
  f\bigl(x,f(f(z,x),x)\bigr)=f(z,x).
\]
Substitution yields $f(x,z)=f(z,x)$ for all $x,z$.
Now fix $x,y$ and put $u=f(x,y)=f(y,x)$.  The third identity gives $f(u,u)=x$, whereas the fourth gives
$f(u,u)=y$.  Hence $x=y$ for every pair of elements, contradicting $|\A|>1$.
\end{proof}

\paragraph{Normalisation, diversification, and the dependency graph.}
Even though Lemma~\ref{lem:sols-strong-unsat} rules out perfect satisfaction, the SDOS identities define a genuine
term-coding optimisation problem: for each $n$ one may ask how many pairs $(x,y)\in\A^2$ can satisfy
\eqref{eq:sols-strong} under an optimal choice of $f$.
Normalising and diversifying is not just technical here; it exposes the dependency graph behind entropy bounds.
For instance, introducing $z:=f(x,y)$ and $w:=f(y,x)$ flattens \eqref{eq:sols-strong} to
\[
  f(x,y)=z,\ \ f(y,x)=w,\ \ f(z,y)=x,\ \ f(x,w)=y,\ \ f(z,w)=x,\ \ f(w,z)=y,
\]
and diversification then replaces the six occurrences of $f$ by $f_1,\dots,f_6$.
The resulting dependency graph has four vertices $\{x,y,z,w\}$.
By Proposition~\ref{prop:div-upper} this unlabelled dependency digraph yields a universal guessing-game (entropy)
upper bound on the maximum code size.
The system is \emph{not} functional in the sense of Definition~\ref{def:fnf}: both $x$ and $y$ appear as outputs of
two distinct equations, so the unlabelled sandwich theorem (Proposition~\ref{prop:sandwich-unlabelled}) is not the right tool
(see the remark after Proposition~\ref{prop:div-upper}).
Instead, the labelled-vertex version (Theorem~\ref{thm:sandwich-labelled}) applies directly: one works with the
labelled dependency graph of Definition~\ref{def:labelled-dep-graph}, in which the two vertices constraining $x$
(and similarly $y$) share a common label and hence enforce a single underlying variable value.

\paragraph{Formulation sensitivity: two scoring conventions for SDOS.}
One subtlety in term coding is that logically equivalent identity systems can yield different optimisation problems,
depending on which variables are treated as \emph{free}.
For SDOS one can either keep $(x,y)$ as the free variables (ideal $n^2$)---this is exactly the scoring problem
induced by \eqref{eq:sols-strong}---or one can rename occurrences to obtain a system with $8$ free variables (ideal
$n^8$).

\smallskip
\noindent\textbf{Formulation 1 ($k=2$ free variables).}
This is the instance with free variables $(x,y)$ and equations \eqref{eq:sols-strong}, whose ideal score is $n^2$.

\smallskip
\noindent\textbf{Formulation 2 ($k=8$ free variables).}
Let $\Gamma_{\mathrm{SDOS}}^{(8)}$ be the instance over the same single binary symbol $f$ with free variables
\(x_1,y_1,\dots,x_4,y_4\) and equations
\begin{equation}
\label{eq:sols-strong-8vars}
\begin{aligned}
& f\bigl(f(x_1,y_1),y_1\bigr)=x_1, \qquad
&& f\bigl(x_2,f(y_2,x_2)\bigr)=y_2,\\[-2pt]
& f\bigl(f(x_3,y_3),f(y_3,x_3)\bigr)=x_3,\qquad
&& f\bigl(f(y_4,x_4),f(x_4,y_4)\bigr)=y_4.
\end{aligned}
\end{equation}
As a universal identity system, \eqref{eq:sols-strong-8vars} is equivalent to \eqref{eq:sols-strong}:
universal validity of \eqref{eq:sols-strong} gives each separated identity, while the converse follows by
substituting $x_1=\cdots=x_4=x$ and $y_1=\cdots=y_4=y$.
As a term-coding optimisation problem, however, it has a different scale ($n^8$ rather than $n^2$) and, as the next
proposition shows, a different exponent.

Geometrically, Formulation~1 counts satisfiable cells in the $n\times n$ grid $\A^2$, whereas Formulation~2 counts
satisfiable points in the four-fold Cartesian product $(\A^2)^4=\A^8$.

\begin{proposition}[Presentation dependence of the SDOS exponent]
\label{prop:sdos-exponents}
Let $\Gamma_{\mathrm{SDOS}}^{(2)}$ denote Formulation~1 and let
$\Gamma_{\mathrm{SDOS}}^{(8)}$ denote Formulation~2.  Then
\[
  \alpha(\Gamma_{\mathrm{SDOS}}^{(2)})=2,
  \qquad
  \alpha(\Gamma_{\mathrm{SDOS}}^{(8)})=8.
\]
Consequently, logical equivalence as universal identity systems does not imply equality of term-coding exponents.
\end{proposition}

\begin{proof}
For Formulation~1 use the flattened system displayed above and diversify its six equations.  A winning
configuration $(x,y,z,w)$ is determined by $(x,y)$, because the first two local rules determine $z$ and $w$.
Hence the labelled game has at most $n^2$ winning configurations.  Equality is attained for every $n$ by choosing
the six local rules so that
\[
  z=x,\qquad w=y,\qquad
  x=z\ \text{from both }(z,y)\text{ and }(z,w),\qquad
  y=w\ \text{from both }(x,w)\text{ and }(w,z).
\]
Thus its diversified labelled game has exactly $n^2$ winning configurations and guessing number $2$.

For Formulation~2, flatten each of the four separated identities and diversify every occurrence of $f$.
All auxiliary variables are determined by the eight original variables, so there are at most $n^8$ winning
configurations.  This bound is attained by projections: in the first component set the inner value to $x_1$ and
decode $x_1$ from it; in the second set the inner value to $y_2$ and decode $y_2$; in the third set the two inner
values to $(x_3,y_3)$ and decode $x_3$ from the first; and in the fourth set them to $(y_4,x_4)$ and decode $y_4$
from the first.  Every assignment to $x_1,y_1,\dots,x_4,y_4$ then extends uniquely to a winning configuration.
The diversified labelled game therefore has exactly $n^8$ winning configurations and guessing number $8$.

Corollary~\ref{cor:exponent} transfers these two labelled guessing numbers to the corresponding term-coding
exponents.
\end{proof}

\subsection{A network-coding micro-example: two-source redundancy}
\label{subsec:network-example}

Term equations capture deterministic network coding directly.
Consider three stored values $(x,y,z)$ where $z$ is computed from $(x,y)$ and each of $(x,z)$ and $(y,z)$
allows recovery of the missing symbol.
Introduce function symbols $f,h_1,h_2:\A^2\to\A$ and impose
\begin{equation}
\label{eq:network}
  z=f(x,y),\qquad h_1(x,z)=y,\qquad h_2(y,z)=x.
\end{equation}
For any $n$ one can interpret $f(x,y)=x+y\pmod n$ and $h_1(u,v)=v-u$, $h_2(u,v)=v-u$, giving $n^2$ valid
triples $(x,y,z)$.
In this example the ideal code size is $n^2$ (two free sources) and is attainable for all $n$.
The dependency graph viewpoint matches the usual entropy calculation: the equations enforce that $(x,y)$
determines $z$ and that $(x,z)$ determines $y$, hence the joint entropy is at most $2$.

\section{Dispersion as a restricted term-coding family}
\label{sec:dispersion}

The main body of this paper focuses on general term-equation systems, where exponents can be non-integers.
A particularly important restricted family in information flow is \emph{dispersion}, where one maximises the
number of distinct output tuples of a term-defined map.
We briefly record how dispersion can be encoded as term coding.

\subsection{Dispersion as an extremal image problem}

Fix terms $t_1(\mathbf x),\dots,t_s(\mathbf x)$ in variables $\mathbf x=(x_1,\dots,x_k)$.
For an interpretation $\mathcal I$ on an alphabet $\A$ of size $n$, consider the induced map
\[
T^{\mathcal I}:\A^k\to \A^s,\qquad
T^{\mathcal I}(\mathbf a):=\bigl(t_1^{\mathcal I}(\mathbf a),\dots,t_s^{\mathcal I}(\mathbf a)\bigr).
\]

\begin{definition}[Dispersion]
\label{def:dispersion}
The \emph{dispersion} of $(t_1,\dots,t_s)$ over alphabet size $n$ is
\[
\mathrm{Disp}_n(t_1,\dots,t_s):=\max_{\mathcal I}\abs{\mathrm{Im}(T^{\mathcal I})}.
\]
\end{definition}

\subsection{Encoding dispersion as Term Coding}

Dispersion counts \emph{distinct output tuples}. Term Coding counts \emph{solutions} (assignments to all
variables). A simple sectioning construction converts dispersion to Term Coding by adding decoding functions that select one
representative preimage for each output.

\begin{proposition}[Dispersion-to-Term-Coding reduction]
\label{prop:disp-to-tc}
Introduce fresh variables $y_1,\dots,y_s$ and fresh function symbols $h_1,\dots,h_k$ of arity $s$.
Let $\Gamma_T$ be the system
\[
  y_i=t_i(\mathbf x)\quad (i=1,\dots,s),
  \qquad\text{and}\qquad
  x_j=h_j(y_1,\dots,y_s)\quad (j=1,\dots,k).
\]
Then for every $n$,
\[\Sn(\Gamma_T)=\mathrm{Disp}_n(t_1,\dots,t_s).\]
\end{proposition}

\begin{proof}
Fix an alphabet $\A$ of size $n$.

\smallskip
\noindent\emph{Upper bound.}
Fix an interpretation $\mathcal I'$ of the enlarged signature and let $\mathcal I$ be its restriction to the
symbols in the $t_i$.
If $(\mathbf x,\mathbf y)$ is a solution of $\Gamma_T$ under $\mathcal I'$, then $\mathbf y=T^{\mathcal I}(\mathbf x)$,
so $\mathbf y\in \mathrm{Im}(T^{\mathcal I})$.
Moreover $\mathbf x$ is uniquely determined by $\mathbf y$ via the equations $x_j=h_j(\mathbf y)$.
Hence the number of solutions is at most $\abs{\mathrm{Im}(T^{\mathcal I})}$, so maximising gives
$\Sn(\Gamma_T)\le \mathrm{Disp}_n(t_1,\dots,t_s)$.

\smallskip
\noindent\emph{Lower bound.}
Fix an interpretation $\mathcal I$ attaining
$\abs{\mathrm{Im}(T^{\mathcal I})}=\mathrm{Disp}_n(t_1,\dots,t_s)$ and write $Y=\mathrm{Im}(T^{\mathcal I})$.
Choose a section $\sigma:Y\to\A^k$ with $T^{\mathcal I}(\sigma(\mathbf y))=\mathbf y$.
Extend $\mathcal I$ to $\mathcal I'$ by setting $h_j^{\mathcal I'}(\mathbf y)=(\sigma(\mathbf y))_j$ for $\mathbf y\in Y$.
Then solutions of $\Gamma_T$ are in bijection with $Y$, so $\Sn(\Gamma_T)\ge \abs{Y}$.
\end{proof}

\begin{remark}
In \cite{riis2019max} dispersion exponents for network-structured instances are characterised by max-flow/min-cut
(and are therefore integers). Proposition~\ref{prop:disp-to-tc} places that theory inside term coding.
In the present paper we emphasise the complementary phenomenon: for general term-equation systems, the exponent
$\alpha(\Gamma)$ can be non-integer.
\end{remark}

\section{Conclusion and further directions}
\label{sec:conclusion}

Term Coding is a deliberately small fragment of equational logic, but maximising the number of satisfying
assignments turns it into a rich extremal theory.
The normalisation and diversification procedures reduce every term-coding instance to a labelled guessing game and
give the two-sided sandwich of Theorem~\ref{thm:sandwich-labelled}.  In functional normal form the labels are
unnecessary, giving the ordinary dependency-digraph sandwich of Proposition~\ref{prop:sandwich-unlabelled} and the
exponent identity $\alpha(\Gamma)=\gn(G_\Gamma,\Src(\Gamma^{\mathrm{nf}}))$.
From an information-theoretic viewpoint, $\log_n \Sn(\Gamma)$ is a normalised entropy maximum under functional
constraints, and the relevant upper bounds live in the geometry of polymatroids and entropy cones.
From a computational viewpoint, the result converts a global optimisation over interpretations
of the function symbols into a local-rule problem on a finite labelled dependency structure,
with an explicit finite-alphabet loss and an exact asymptotic rate.

Several directions look particularly natural.
\begin{enumerate}[leftmargin=2.2em]
\item \textbf{Rates of convergence.}  Theorem~\ref{thm:gn-converges} shows that for every fixed (labelled) digraph the
  guessing number $\gn(G,\lambda,S,n)$ converges to its limit $\gn(G,\lambda,S)$ as $n\to\infty$.
  The proof gives the explicit one-sided bound \eqref{eq:gn-floorlog}: for every fixed $q$,
  $\gn(\cdot,n)\ge\gn(\cdot,q)\bigl(1-1/(\lfloor\log_q n\rfloor+1)\bigr)$, a loss of order
  $1/\log n$ relative to $\gn(\cdot,q)$, but no rate of approach to the supremum itself.  How
  fast can the convergence be?  Even coarse quantitative bounds would make the sandwich theorem effective at
  finite $n$.
\item \textbf{Presentation dependence.}  Logically equivalent universal identity systems can define genuinely
  different optimisation problems, including different exponents, as Proposition~\ref{prop:sdos-exponents} shows.
  A systematic theory of which presentation changes preserve the exponent would be useful.
\item \textbf{Beyond pure equations.}  Two extensions that remain close in spirit are (i) adding consistent
  disequality constraints, and (ii) passing to multiple sorts.  Both appear naturally in coding and finite
  model theory and can be incorporated without changing the basic graph/entropy mechanism.
\item \textbf{Algorithmic questions.}  Computing $\gn(G)$ exactly is difficult in general.  It would be
  interesting to identify robust graph classes (beyond the max-flow cases from dispersion) for which the
  exponent, or good approximations to it, are efficiently computable.
\end{enumerate}

\section*{Acknowledgements}
This paper grew out of earlier work on guessing numbers, entropy methods and information flow.
The core mathematical ideas---the term-coding framework, the guessing-number sandwich theorem,
and the principal examples---were developed by the author, with AI tools used only for editing,
and are documented in the first arXiv version, arXiv:2601.16614v1 (January 2026).
In preparing the present version, language-model assistants were used as tools for editing and
shortening the manuscript, drafting the accompanying Lean~4 development, and conducting
a separate statement-by-statement audit of the manuscript against that development.
This process led, in particular, to the precise formulation and proof of
Proposition~\ref{prop:c5-vertex} and to several clarifying remarks.
The author has checked all results and is responsible for the entire content.

% Print bibliography with a standard heading (helps arXiv/journal builds)
\begingroup
\emergencystretch=1em
\printbibliography
\endgroup

\end{document}